\documentclass[12pt]{article}
\usepackage{booktabs}  % professionally typeset tables
\usepackage{amsmath}
\usepackage{amsthm}
\usepackage{amssymb}
\usepackage{mathrsfs}
\usepackage{gensymb}
\usepackage{textcomp}  % better copyright sign, among other things
\usepackage{xcolor}
\usepackage{lipsum}    % filler text
\usepackage{graphicx}
\usepackage{caption}
\usepackage{subcaption}
\usepackage{bm, bbm}
\RequirePackage{diagbox, adjustbox}
\usepackage{natbib}
\usepackage[colorlinks=true,linkcolor=black, citecolor=black, urlcolor=black]{hyperref}
\usepackage{multirow}
\usepackage{multicol}
\usepackage{pdflscape}
\usepackage{afterpage}
\usepackage{everypage}
\usepackage{natbib}
\usepackage{url} % not crucial - just used below for the URL 
 
\newcommand{\norm}[1]{\left\lVert#1\right\rVert}
\DeclareMathOperator*{\argminB}{argmin}
\newcommand{\blind}{1}

\begin{document}

\def\spacingset#1{\renewcommand{\baselinestretch}%
{#1}\small\normalsize} \spacingset{1}

%%%%%%%%%%%%%%%%%%%%%%%%%%%%%%%%%%%%%%%%%%%%%%%%%%%%%%%%%%%%%%%%%%%%%%%%%%%%%%

\if1\blind
{
  \title{\bf Estimating Atmospheric Motion Winds from Satellite Image Data using Space-time Drift Models}
  \author{Indranil Sahoo$^1$, Joseph Guinness$^2$, and Brian J. Reich$^3$ \\ \\
    $^1$Department of Statistical Sciences and Operations Research, \\ Virginia Commonwealth University\\ 
    $^2$Department of Statistics and Data Science, Cornell University \\
    $^3$Department of Statistics, North Carolina State University
    }
    \date{}
  \maketitle
} \fi

\if0\blind
{
  \bigskip
  \bigskip
  \bigskip
  \begin{center}
   {\LARGE\bf Estimating Atmospheric Motion Winds from Satellite Image Data using Space-time Drift Models}
\end{center}
 \medskip
} \fi

%\bigskip
\begin{abstract}
Geostationary weather satellites collect high-resolution data comprising a series of images. The Derived Motion Winds (DMW) Algorithm is commonly used to process these data and estimate atmospheric winds by tracking features in the images. However, the wind estimates from the DMW Algorithm are often missing and do not come with uncertainty measures. This motivates us to statistically model wind motions as a spatial process drifting in time. Using a covariance function that depends on spatial and temporal lags and a drift parameter to capture the wind speed and wind direction, we estimate the parameters by local maximum likelihood. Our method allows us to compute standard errors of the local estimates, enabling spatial smoothing of the estimates using a Gaussian kernel weighted by the inverses of the estimated variances. We conduct extensive simulation studies to determine the situations where our method performs well. The proposed method is applied to the GOES-15 brightness temperature data over Colorado and reduces prediction error of brightness temperature compared to the DMW Algorithm. 

\end{abstract}

\noindent%
\vfill

\newpage
\spacingset{1.45} % DON'T change the spacing!

\section{Introduction}

Wind field estimates are key inputs to many important environmental studies.  For example, winds are an important component of atmospheric circulation and thus estimating local or regional winds is important for making weather forecasts. Wind also influences vegetation as they affect factors that influence plant growth, such as seed dispersal rates, air transportation of pollen, and metabolism rates in plants. The study of local winds also permits evaluation of power produced by wind turbines \citep{brown1984time,castino1998stochastic}, prediction of propagation of oil-spills \citep{kim2014analysis} and the study of coastal erosion \citep{ahmad2015analyses}. 
Local winds are capable of dispersive air pollutants, such as the local weather phenomenon known as Calima that blows dust into the Canary Islands \citep{weatheronline}. Winds may also impact large-scale devastation such as forest fires. For example, the Santa Ana winds blow into California after scorching summers \citep{wildfires} and produce conditions that lead to large wildland fires. Thus, it is important to map the strength and direction of local winds to help us prepare for natural calamities and facilitate preventive measures.

Observations of wind speed and direction at the ground level are collected at weather stations on land and by buoys or ships over oceans.
There is a significant statistical literature on modeling winds from ground monitors \citep{priestley1981spectral, haslett1989space, brillinger2001time, stein2005}. For some applications, it is sufficient to analyze the evolution of wind at a fixed location, and several methods have been proposed for this scenario \citep{brown1984time,tol1997autoregressive,ailliot2004modeles, monbet2007survey}. The wind at different locations can be utilized to model spatio-temporal dependencies \citep{bennett1979spatial, bras1985random, kyriakidis1999geostatistical, de2005predictive}. Also, \citet{boukhanovsky2003stochastic}, \citet{malmberg2005forecasting} and \citet{ailliot2006autoregressive} proposed autoregressive space-time models to describe the evolution of winds. \citet{stein2005} proposed a spectral-in-time modeling approach to describe the space-time dependencies of the data. \citet{fuentes2008spatial} modeled a drift process using Bayesian analysis, where the drift parameter is modeled using splines. \citet{modlin2012circular} used circular conditional autoregressive models for wind direction and speed. \citet{sigrist2012dynamic} modeled precipitation with
a non-separable spatio-temporal model using an external wind vector. However, these wind observations have very sparse coverage.  

Low orbit satellites such as Jason 3 and Sentinel 1 infer surface winds using geophysical inversion algorithms, based on peak backscattered power and the shape of radio signal waveforms \citep{sentinelonline}. However, ground monitors are sparse in space, and the satellites cannot monitor winds continuously in space and time, as they are in low earth orbit. Winds in the upper level of the atmosphere can be observed using weather balloons or aircraft measurements, but these observations are also very sparse in space and time.

\begin{figure}
\centering
\begin{minipage}{0.32\textwidth}
\includegraphics[width=\linewidth]{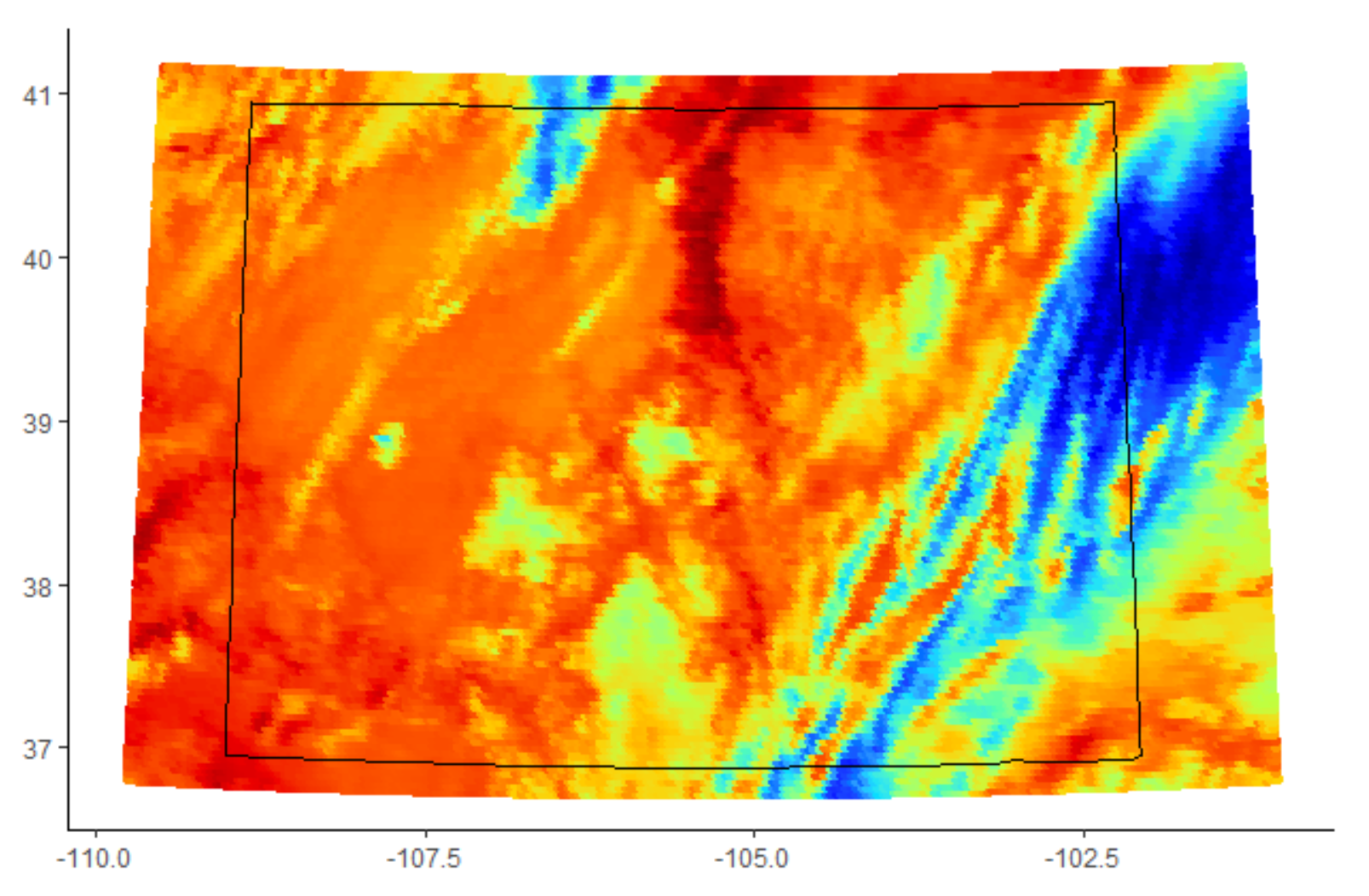}
\end{minipage} %
\begin{minipage}{0.32\linewidth}
\includegraphics[width=\linewidth]{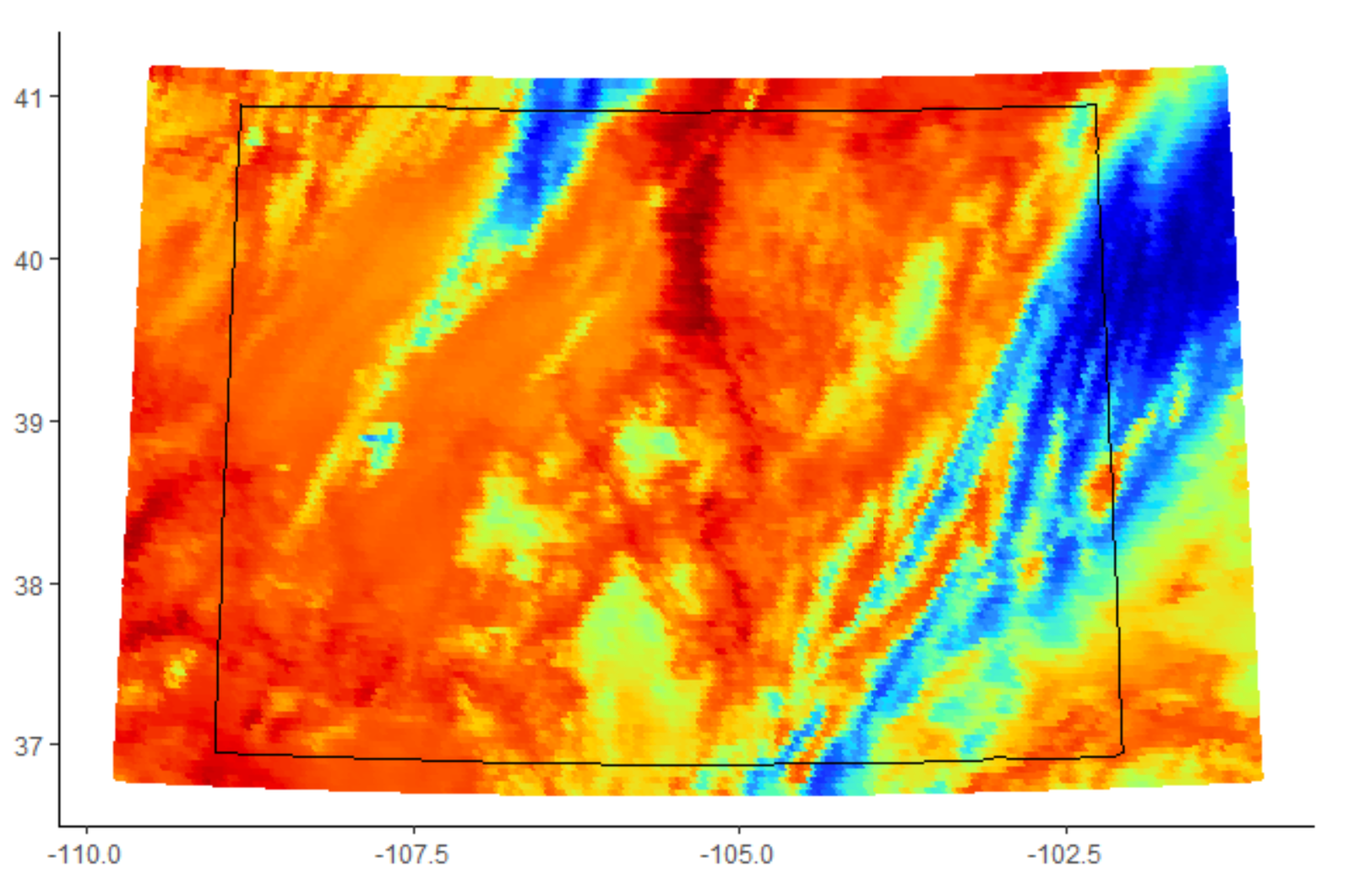}
\end{minipage} %
\begin{minipage}{0.345\linewidth}
\includegraphics[width=\linewidth]{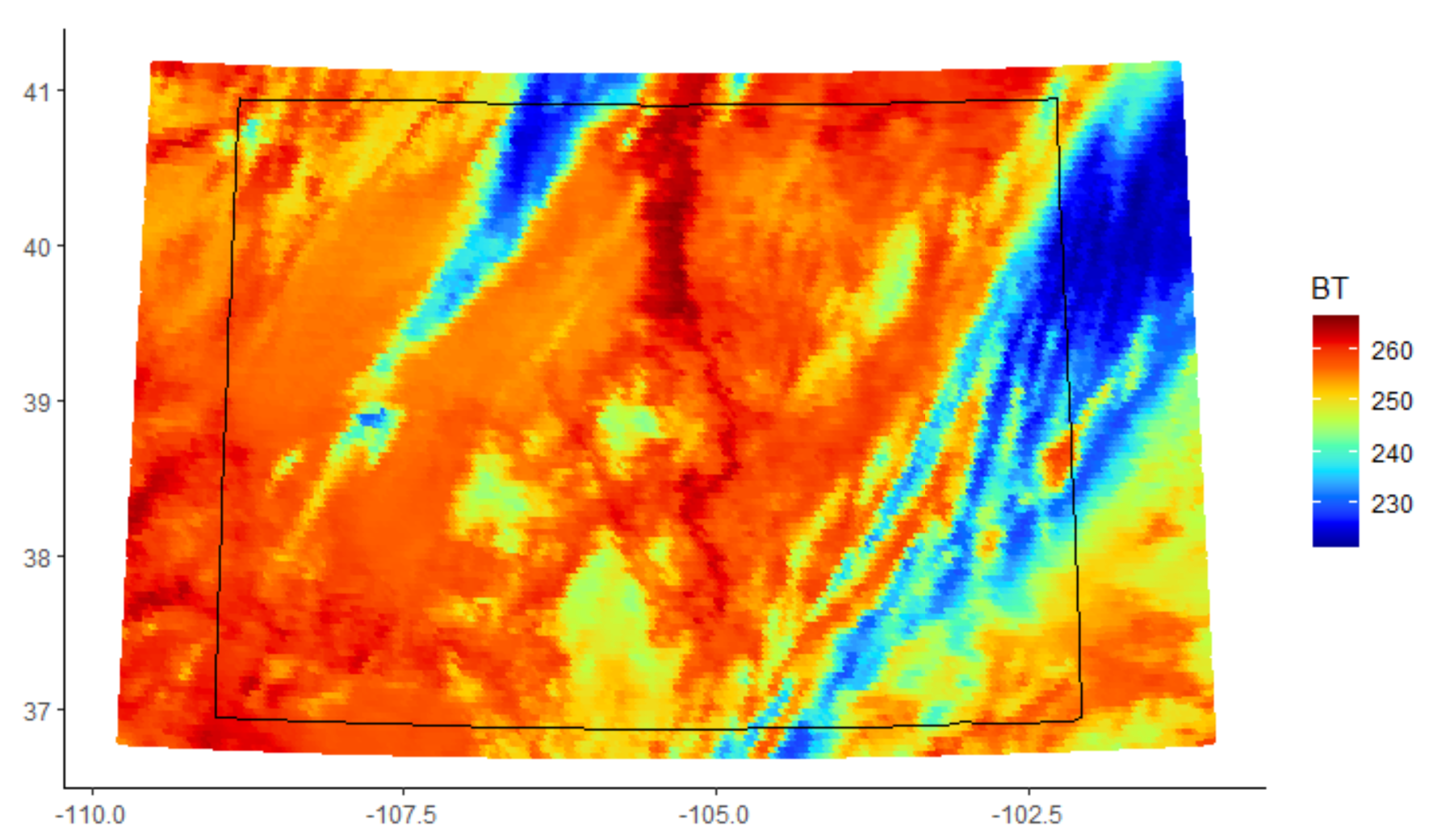}
\end{minipage}
\caption{A sequence of Brightness temperature (Kelvin) images over Colorado, captured by the GOES-15 satellite on January 3, 2015 at three consecutive time points.}
\label{fig1a}
\end{figure}

On the other hand, geostationary weather satellites provide data from the surface and the atmosphere with a very high temporal resolution. The resulting data comprise a series of images which essentially make them a `movie'. While the satellites do not directly measure wind, they measure  infrared channel radiations in terms of brightness temperature. These brightness temperature image sequences are used to infer wind estimates by tracking movements of atmospheric tracers such as clouds or moisture features over time. Figure \ref{fig1a} shows a sequence of brightness temperature images captured by the Geostationary Operational Environmental Satellite (GOES) - 15, operated by the National Oceanic and Atmospheric Administration (NOAA). Wind data obtained from satellite images play a major role in data assimilation \citep{lahoz2014data}. Numerical climate models perform better with accurate wind data, especially over the oceans, resulting in improved weather forecasts and warnings \citep{tomassini1999use}. For example, the European Centre for Medium-Range Weather Forecasts (ECMWF) has been incorporating atmospheric motion winds into their forecast models operationally since the 1980s. This has dramatically improved the model's ability to forecast the track of tropical cyclones and has also increased the model's ability to predict wave heights and storm surges \citep{tomassini1999use}. 

The Derived Motion Winds (DMW) Algorithm \citep{daniels2010goes} is a standard algorithm for estimating motion winds from satellite images. The DMW Algorithm takes as input brightness temperature images from the National Oceanic and Atmospheric Administration (NOAA) geostationary meteorological satellites and gives estimated wind fields as outputs. The algorithm tracks a suitable target across the input images and assigns a motion wind to the middle time point (see Section 2 for details). The image at the middle time point must satisfy a set of criteria to qualify as a suitable target scene. As a result, the DMW estimates are often sparse. The DMW Algorithm also does not provide a measure of uncertainty. 

Geostatistical methods are capable of overcoming these shortcomings. This motivates us to model satellite image data using a spatial process drifting in time. At the heart of this statistical model lies the idea of incorporating the motion vector parameters in the process covariance. \citet{stein2013stochastic} uses this idea to fit a space-time model. We borrow the idea of Nested Tracking from \citet{daniels2010goes} by considering data buffers and estimate local wind vectors using maximum likelihood estimates over sliding windows over space. Local estimation of covariance parameters using moving windows has been studied in \citet{haas1990lognormal,haas1995local, risser2015local, ver2004flexible}.  Estimating covariance parameters locally helps tackle the non-stationarity present in the data. One major advantage of our approach is that it circumvents the computational burden for large sample sizes. Moreover, the proposed algorithm can easily be run in parallel, which can lead to further computational improvements. Another major advantage of the approach over the DMW Algorithm is that it allows us to quantify uncertainties in the local estimates of the winds; this in turn facilitates our use of an inverse-variance-weighted spatial smoothing algorithm to borrow strength across space while down-weighting the most uncertain estimates. This also allows trade offs between bias and variance for reducing estimation error \citep{anderes2011local}.

\section{Derived Motion Winds Algorithm}

The Derived Motion Winds (DMW) Algorithm estimates atmospheric motion winds from images taken by  geostationary satellites. For a cloudy region, the imager records brightness temperature (see Figure \ref{fig2a}), which measures the radiance (in Kelvin) of microwave radiation traveling upward from the top of the atmosphere to the satellite. For clear sky portions, the satellite records images of suitable indicators of atmospheric moisture content, such as specific humidity. \citet{daniels2010goes} provides a description of and the physical basis for the estimation of atmospheric winds from the images taken by geostationary satellites.  

The DMW Algorithm involves creating a \textit{data buffer}, which is a data structure holding 2-dimensional arrays of brightness temperature for 3 consecutive image times. The middle portion of the buffer is divided into smaller target scenes (or target windows), and each scene is analyzed to locate and select a set of suitable targets in the middle image. 

\citet{daniels2010goes} also gives a description of the Nested Tracking Algorithm which involves nesting smaller target windows (usually of size $5 \times 5$) within a large target scene of size $15 \times 15$ pixels and getting every possible local motion vectors derived from each possible smaller window within a large target scene. The displacement vector between time points $t$ and $t + 1$ is computed by minimizing the Sum of Squared Differences (SSD) criterion as
$$
\widehat{\bm{v}}(\bm{x}, t, t + 1) = \argminB_{\bm{u}} \sum_{\bm{s} \in D_{\bm{x}}} \lbrace Y(\bm{s}, t) - Y(\bm{s} + \bm{u}, t + 1)\rbrace ^2,
$$
where $Y(\bm{s}, t)$ denotes the brightness temperature within the smaller window at pixel location $\bm{s}$ and time point $t$, $D_{\bm{x}}$ is the indices of pixels in the target window centered at $\bm{x}$, and $\bm{u}$ is a two-dimensional vector denoting the displacement. The sum is considered over two dimensions and the optimization over $\bm{u}$ is done only over integers so that $\bm{s} + \bm{u}$ corresponds to an observed pixel. In practice, the target scene is substantially larger than the size of the smaller target window, so the above summation is carried out for all target window positions within the target scene. The mean displacement vector is computed as 
$$
\widehat{\bm{u}}(\bm{x}, t) = \frac{1}{2}\lbrace \widehat{\bm{v}}(\bm{x}, t - 1, t) + \widehat{\bm{v}}(\bm{x}, t, t + 1)\rbrace
$$
and is assigned as the DMW estimate at location $\bm{x}$ time point $t$ in the buffer. Once every possible local motion vector within the buffer are calculated, a density-based cluster analysis algorithm, DBSCAN \citep{ester1996density} is used to identify the largest cluster representing the dominant motion. The final DMW estimate for the buffer is the average of the vectors belonging to the largest cluster. 

The size of the target window depends on the spatial and temporal resolution of the imagery and the scale of the intended feature to be tracked. \citet{daniels2010goes} suggests that the temporal resolution of the images should at most be 15 minutes in order to account for the short lifespan and rapid disintegration of clouds over land. The DMW Algorithm does not offer wind estimates at every space-time location as the data in the middle image has to satisfy a set of criteria to qualify as a suitable target scene \citep{daniels2010goes}. Also, quantifying uncertainties using the SSD criterion is hard because each estimated vector uses a different subset of the data, so likelihood ratio tests are not applicable. Finally, the highest resolution of the vector estimates generated for each target scene can be half-integers. 
 
\section{Model-based wind estimation}

\subsection{Space-time drift models}

The proposed approach uses spatio-temporal covariance functions to track the wind. This requires us to consider asymmetric spatio-temporal covariance functions. Space-time asymmetries in covariance functions have been studied in \citet{stein2005space}, \citet{huang2019visualization} and \citet{park2006new}. 

The space-time process $Z(\bm{x},t)$ has asymmetric covariance if
\begin{equation} \label{eq3.1}
\mbox{Cov} \lbrace Z(\bm{x}, t_1), Z(\bm{y}, t_2) \rbrace \neq \mbox{Cov} \lbrace Z(\bm{x}, t_2), Z(\bm{y}, t_1) \rbrace.
\end{equation}
Here $(\bm{x}, t)$ is a space-time location, with $\bm{x} = (x_1, x_2)$ giving the spatial coordinates, and $t$ giving the time. In most regions, winds flow in a consistent direction, and so changes in brightness temperature and other atmospheric variables at one location tend to precede similar changes down wind. For instance, if $t_2 > t_1$ and winds flow consistently from $\bm{x}$ to $\bm{y}$, then  we expect $$\mbox{Cov} \lbrace Z(\bm{x}, t_1), Z(\bm{y}, t_2) \rbrace > \mbox{Cov} \lbrace Z(\bm{x}, t_2), Z(\bm{y}, t_1) \rbrace.$$

We incorporate space-time asymmetries via a drift parameter. Suppose that $Z_0: \mathbb{R}^3 \rightarrow \mathbb{R}$ is a stationary, space-time symmetric process with covariance between $Z_0(\bm{x}, t_1)$ and $Z_0(\bm{y}, t_2)$ given by
$$C_0 \lbrace (\bm{x}, t_1), (\bm{y}, t_2) \rbrace = M_{\nu}\left( \left\{ \frac{[x_1 - y_1 + a_{12}(x_2 - y_2)]^2}{(1/a_{11})^2} + \frac{(x_2 - y_2)^2}{(1/a_{22})^2} + \frac{(t_1 - t_2)^2}{(1/a_{33})^2}\right\}^{1/2} \right).
$$
Here $M_{\nu}$ is the isotropic Mat\'ern \citep{stein2012interpolation} covariance function, 
$$
M_{\nu}(d) = \frac{\sigma^2}{2^{\nu - 1}\Gamma(\nu)}(d)^{\nu}\kappa_{\nu}(d),
$$
where $d = ||\bm{x} - \bm{y}||_2$, $\kappa_{\nu}(\cdot)$ is the modified Bessel
function of the second kind of order $\nu$, and $\Gamma(\cdot)$ is the gamma function. $\sigma^2$ is the spatial process variance, and $\nu$ is the smoothness parameter which controls the mean square differentiability of the process. Also, $a_{11}^2$ and $a_{22}^2$ are the inverse spatial range parameters and $a_{33}^2$ is the inverse temporal range parameter. $Z_0$ is geometrically anisotropic in its spatial coordinates with anisotropy parameter $a_{12}$, but it is space-time symmetric
because 
$$
C_0 \lbrace (x_1, x_2, t_1), (y_1, y_2, t_2) \rbrace = C_0 \lbrace (y_1, y_2, t_1), (x_1, x_2, t_2) \rbrace.
$$
Now, let us define
$$
Z(\bm{x}, t) = Z_0(\bm{x} + \bm{u}t, t),
$$
where the two-dimensional vector $\bm{u} = (u_1, u_2)$ is the drift of the process over time and can be interpreted as the zonal and meridional components of wind. 
Then the covariance function of $Z(\bm{x}, t)$ is
\begin{equation} \label{eq3.2}
\begin{aligned}
 &\mbox{Cov}\lbrace Z(\bm{x}, t_1), Z(\bm{y}, t_2)\rbrace \\
 &= C_0 \lbrace (x_1 + u_1t_1, x_2 + u_2t_1, t_1),  (y_1 + u_1t_2, y_2 + u_2t_2, t_2)\rbrace \\
&= M_{\nu}\left( \left\{ \frac{[x_1 - y_1 + a_{12}(x_2 - y_2) + (u_1 + a_{12}u_2)(t_1 - t_2)]^2}{(1/a_{11})^2} + \frac{[(x_2 - y_2) + u_2(t_1 - t_2)]^2}{(1/a_{22})^2} + \frac{(t_1 - t_2)^2}{(1/a_{33})^2}\right\}^{1/2} \right)  \\
\end{aligned}
\end{equation}
which is space-time asymmetric and stationary.

\subsection{Local estimation of the drift parameter}

Assume the brightness temperature data $Y(\bm{x},t)$ are normally distributed and have been standardized at each location (as described in the Appendix); denote the standardized data as $Z(\bm{x},t)$.  The mean and variance carry no information about the drift, and this step simplifies estimation of the correlation parameters. We do not specify a global covariance function for $Z$. Instead, we specify its local covariance with drift parameter $\bm{u}(\bm{x}, t) \in \mathbb{R}^2$ and estimate the motion winds locally. 

We define a target window as a square array of pixels
$$
D(\bm{x}, t) = \lbrace (\bm{x}', t'): \norm{\bm{x} - \bm{x}'}_{\infty} < \epsilon \text{ \& } |t - t'| \leq 1 \rbrace .
$$
To estimate $\bm{u}(\bm{x}, t)$, we assume that the process $Z(\bm{x}, t)$ is locally stationary in $D(\bm{x}, t)$ and that winds are smooth enough to be assumed constant in the scene, that is, $\bm{u}(\bm{x}', t') \approx \bm{u}(\bm{x}, t) \text{ for all } (\bm{x}', t') \in D(\bm{x}, t)$. In other words, we approximate the local covariance function as  
$$
\mbox{Cov} \lbrace Z(\bm{x}_1,t_1), Z(\bm{x}_2, t_2) \rbrace \approx C_0 \lbrace (\bm{x}_1 + \bm{u}(\bm{x}, t)t_1, t_1),  (\bm{x}_2 + \bm{u}(\bm{x}, t)t_2, t_2)\rbrace
$$
for $(\bm{x}_1, t_1) \text{ and } (\bm{x}_2, t_2) \in D(\bm{x}, t)$. 

We use maximum likelihood estimation in local, moving windows to make inference about the spatially varying covariance parameters $\bm{\theta}(\bm{x}, t) = (a_{11}^2(\bm{x}, t), a_{22}^2(\bm{x}, t), a_{33}^2(\bm{x}, t), a_{12}(\bm{x}, t), \bm{u}(\bm{x}, t))$. The local estimation approach deals with the non-stationarity present in the entire data. We use maximum likelihood estimation within $D(\bm{x}, t)$ to estimate $\bm{\theta}(\bm{x}, t) \equiv \bm{\theta}_D$. If $\bm{Z}_D$ denotes the standardized data vector in the target window and $\Sigma(\bm{\theta}_D)$ denote the corresponding space-time covariance matrix with elements defined by \eqref{eq3.2}, then the log-likelihood for $ \bm{\theta}_D$ given $\bm{Z}_D$ is 
\begin{equation}\label{eq3.4}
l\left(\bm{\theta}_D|\bm{Z}_D\right) = -\frac{1}{2}\mbox{log}\left(|\Sigma(\bm{\theta}_D)|\right) - \frac{1}{2} \bm{Z}_D^T \lbrace \Sigma(\bm{\theta}_D) \rbrace ^{-1}\bm{Z}_D.
\end{equation}

The estimated wind vectors are associated with the space-time location ($\bm{x}$, t) at which the target window $D$ was centered, denoted $\widehat{\bm{u}}(\bm{x}, t)$. We also estimate the variances associated with the estimated wind vectors by computing the Fisher information at the maximum likelihood estimate.  We imitate the Nested Tracking approach and slide the target window across space and time, estimating wind vectors locally in space and time using the same optimization routine. In Section 4.1, we use the exact likelihood in (\ref{eq3.4}) to estimate the wind vectors. However, in Sections 4.2 and 5, we use a fast Vecchia approximation of the likelihood function \citep{guinness2018permutation}, implemented in the `GpGp' R package \citep{guinness2018gpgp}.

\subsection{Smoothing the local estimates}
After obtaining the local estimates of $\bm{u}(\bm{x},t)$ for all $(\bm{x},t)$, we smooth these initial estimates to stabilize them by borrowing strength across space.  The two components of the wind vectors are smoothed separately.  The kernel smoothing weights are taken to be proportional to the ratio of a spatial Gaussian kernel and the variance of the initial estimate.  Full details are given in the Appendix.  The bandwidth is chosen based on cross validation. 

The size of the target window is an important tuning parameter. While implementing the method on real data sets, the window size is chosen such that the wind motion is roughly constant in the scene, and the feature being tracked in time is prominent and does not move out of frame. In Section 4, we perform a simulation study analyzing the effect of the model parameters including window size on the performance of our method. We also analyze the efficacy of spatial smoothing of the local estimates. While implementing the methods in the real-data analysis in Section 5, the optimal window size is chosen using cross validation. This works well as wind fields are mostly smooth over a fairly small region and time frame.

\section{Simulation studies}

\subsection{Motion wind estimation}

In the first part of our simulation study, we determine the conditions under which the space-time drift model (STDM) performs well for estimating motion winds. For this purpose, we repeatedly simulate data sets within one particular target window (as opposed to scanning across a spatial domain) from a Gaussian process with exponential covariance (that is, $M_{\nu}$ with $\nu = 1/2$ and $a_{12} = 0$). We also implement a version of the DMW Algorithm and compare its performance with the STDM. To compare the two methods, accuracy for simulated data set $i$ is measured by Vector Difference \citep{daniels2010goes} between the true ($\bm{u}_0$) and estimated ($\widehat{\bm{u}}_i$) wind vectors
\begin{align*}
\mbox{VD}_i &= \norm{\widehat{\bm{u}}_i - \bm{u}_0} , 
\end{align*}
and we report the mean and standard deviation of $\mbox{VD}_1, \ldots, \mbox{VD}_N$ over $N = 100$ data sets in Tables \ref{3t1} - \ref{3t3}. 

First we consider repeated data simulations in a target window of size $7 \times 7 \times 3$, generated independently from a Gaussian process with covariance function $M_{\nu}$ with $\nu = 1/2$ and $\bm{u}(\bm{x}, t) = \bm{u}_0$ (constant true wind) using a fast Gaussian approximation algorithm \citep{guinness2018permutation} which is implemented using the `GpGp' package \citep{guinness2018gpgp} in R. We assume $1/a_{11}^2 = 1/a_{22}^2 = \alpha_1^2$ and the true spatial range parameter, $\alpha_1^2$ is chosen to be either 1, 2, 4 or 8. The true temporal range parameter $1/a_{33}^2 = \alpha_2^2$ is chosen to be either 1, 2, 3 or 4. We also take two different values of the true wind vector, namely $\bm{u}_0 = (1, 2)^T$ and $(3, 5)^T$, which signify respectively slow and fast wind vectors. All four parameters are updated simultaneously during optimization. Table \ref{3t1} shows the performance of the two methods for the two wind vectors based on $N = 100$ simulations. 

\begin{table}
\vspace*{1 cm}
\caption{\label{3t1}Comparing Mean Vector Difference (SD) for Space-time Drift Model (STDM) and Derived Motion Winds algorithm (DMWA) for true wind vectors $\bm{u}_0 = (1, 2)^T$ (left panel) and $\bm{u}_0 = (3, 5)^T$ (right panel) based on data window of size $7 \times 7$; $\alpha_1^2$ and $\alpha_2^2$ denote the true spatial and temporal range respectively. MVD and SD are measured in pixels.}
\centering
\noindent \resizebox{\textwidth}{!}{%
\begin{tabular}{ |ccccc| }
 \multicolumn{5}{c}{MVD of STDM for $\bm{u}_0 = (1, 2)^T$}   \\ \hline
  \backslashbox{$\alpha_1^2$}{$\alpha_2^2$} & 1 & 2 & 3 & 4 \\ 
  \hline
  1 & 1.196 (0.35) & 0.311 (0.22) & 0.243 (0.24) & 0.201 (0.20) \\ 
  2 & 1.658 (1.13) & 0.600 (0.52) & 0.353 (0.33) & 0.245 (0.12) \\ 
  4 & 2.841 (1.75) & 1.599 (1.24) & 0.861 (0.61) & 0.456 (0.31) \\
  8 & 3.253 (2.25) & 3.281 (1.87) & 2.212 (1.71) & 1.432 (1.02)  \\ \hline
  \multicolumn{5}{c}{MVD of DMWA for $\bm{u}_0 = (1, 2)^T$} \\ \hline
  \backslashbox{$\alpha_1^2$}{$\alpha_2^2$} & 1 & 2 & 3 & 4 \\ 
  \hline
 1 & 1.983 (1.05) & 1.209 (0.85) & 1.039 (1.12) & 0.854 (1.09) \\
 2 & 2.045 (1.02) & 1.527 (0.94) & 1.283 (1.01) & 1.072 (0.98) \\
 4 & 2.342 (0.94) & 2.158 (0.94) & 1.620 (0.92) & 1.511 (0.91) \\
 8 & 2.452 (1.07) & 2.175 (0.95) & 2.003 (1.10) & 1.776 (0.89) \\ \hline
\end{tabular}
\begin{tabular}{ |ccccc| }
 \multicolumn{5}{c}{MVD of STDM for $\bm{u}_0 = (3, 5)^T$}   \\ \hline
  \backslashbox{$\alpha_1^2$}{$\alpha_2^2$} & 1 & 2 & 3 & 4 \\ 
  \hline
  1 & 2.310 (1.60) & 2.034 (1.66) & 1.734 (1.68) & 1.787 (1.95) \\ 
  2 & 2.335 (1.56) & 2.032 (1.76) & 1.392 (1.47) & 1.030 (1.21) \\ 
  4 & 3.440 (2.34) & 2.393 (1.90) & 1.996 (1.70) & 1.638 (1.99) \\
  8 & 3.611 (2.66) & 3.135 (1.92) & 2.913 (1.88) & 2.483 (2.39)  \\ \hline
  \multicolumn{5}{c}{MVD of DMWA for $\bm{u}_0 = (3, 5)^T$} \\ \hline
  \backslashbox{$\alpha_1^2$}{$\alpha_2^2$} & 1 & 2 & 3 & 4 \\ 
  \hline
 1 & 5.953 (0.91) & 5.924 (0.96) & 5.953 (1.10) & 6.012 (0.98) \\
 2 & 5.710 (1.09) & 5.781 (1.18) & 5.899 (1.09) & 5.739 (1.08) \\
 4 & 5.665 (1.09) & 5.577 (1.16) & 5.540 (1.21) & 5.355 (1.13) \\
 8 & 5.952 (1.13) & 5.678 (1.23) & 5.413 (1.12) & 5.461 (1.18) \\ \hline
\end{tabular}}
\end{table}

\begin{table}
\vspace*{1 cm}
\caption{\label{3t2}Comparing Mean Vector Difference (SD) for Space-time Drift Model (STDM) and Derived Motion Winds algorithm (DMWA) for true wind vectors $\bm{u}_0 = (1, 2)^T$ (left panel) and $\bm{u}_0 = (3, 5)^T$ (right panel) based on data window of size $11 \times 11$; $\alpha_1^2$ and $\alpha_2^2$ denote the true spatial and temporal range respectively. MVD and SD are measured in pixels.}
\centering
\noindent \resizebox{\textwidth}{!}{%
\begin{tabular}{ |ccccc| }
\multicolumn{5}{c}{MVD of STDM for $\bm{u}_0 = (1, 2)^T$}   \\ \hline
\backslashbox{$\alpha_1^2$}{$\alpha_2^2$} & 1 & 2 & 3 & 4 \\ \hline
1 & 0.415 (0.37) & 0.172 (0.09) & 0.136 (0.07) & 0.118 (0.05) \\ 
2 & 1.225 (1.21) & 0.304 (0.19) & 0.162 (0.09) & 0.149 (0.08) \\ 
4 & 3.010 (2.56) & 1.008 (0.72) & 0.398 (0.25) & 0.268 (0.14) \\
8 & 3.441 (3.48) & 2.830 (2.06) & 1.346 (1.00) & 0.831 (0.69)  \\ \hline
\multicolumn{5}{c}{MVD of DMWA for $\bm{u}_0 = (1, 2)^T$} \\ \hline
\backslashbox{$\alpha_1^2$}{$\alpha_2^2$} & 1 & 2 & 3 & 4 \\ 
  \hline
1 & 1.965 (1.17) & 0.771 (0.91) & 0.162 (0.50) & 0.084 (0.42) \\
2 & 2.230 (1.30) & 1.343 (1.08) & 0.727 (0.82) & 0.409 (0.69) \\
4 & 2.532 (1.32) & 1.888 (1.18) & 1.856 (1.17) & 1.101 (0.86) \\
8 & 2.891 (1.54) & 2.538 (1.33) & 2.064 (1.09) & 1.961 (1.19) \\ \hline
\end{tabular}
\begin{tabular}{ |ccccc| }
 \multicolumn{5}{c}{MVD of STDM for $\bm{u}_0 = (3, 5)^T$}   \\ \hline
  \backslashbox{$\alpha_1^2$}{$\alpha_2^2$} & 1 & 2 & 3 & 4 \\ 
  \hline
 1 & 1.124 (1.09) & 0.868 (1.50) & 0.743 (1.57) & 0.530 (1.06) \\ 
 2 & 1.916 (1.69) & 0.777 (1.24) & 0.299 (0.42) & 0.230 (0.41) \\ 
 4 & 2.820 (1.98) & 1.489 (1.36) & 0.676 (0.70) & 0.392 (0.37) \\
 8 & 3.401 (2.84) & 3.125 (1.96) & 2.071 (1.80) & 1.129 (1.03)  \\ \hline
  \multicolumn{5}{c}{MVD of DMWA for $\bm{u}_0 = (3, 5)^T$} \\ \hline
  \backslashbox{$\alpha_1^2$}{$\alpha_2^2$} & 1 & 2 & 3 & 4 \\ 
  \hline
 1 & 5.823 (1.62) & 5.836 (1.43) & 5.866 (1.59) & 6.024 (1.61) \\
 2 & 5.638 (1.74) & 5.294 (1.75) & 5.408 (1.80) & 5.362 (1.64) \\
 4 & 5.548 (1.42) & 4.995 (1.65) & 4.934 (1.85) & 4.750 (1.70) \\
 8 & 5.864 (1.62) & 5.320 (1.72) & 5.233 (1.75) & 4.603 (1.70) \\ \hline
\end{tabular}%
}
\end{table}

\vspace{1cm}

\begin{table}
\vspace*{1 cm}
\caption{\label{3t3}Comparing Mean Vector Difference (SD) for Space-time Drift Model (STDM) and Derived Motion Winds algorithm (DMWA) for true wind vectors $\bm{u}_0 = (1, 2)^T$ (left panel) and $\bm{u}_0 = (3, 5)^T$ (right panel) based on data window of size $15 \times 15$; $\alpha_1^2$ and $\alpha_2^2$ denote the true spatial and temporal range respectively. MVD and SD are measured in pixels.}
\centering
\noindent \resizebox{\textwidth}{!}{%
\begin{tabular}{ |ccccc| }
 \multicolumn{5}{c}{MVD of STDM for $\bm{u}_0 = (1, 2)^T$}   \\ \hline
  \backslashbox{$\alpha_1^2$}{$\alpha_2^2$} & 1 & 2 & 3 & 4 \\ 
  \hline
  1 & 0.274 (0.32) & 0.125 (0.06) & 0.097 (0.05) & 0.073 (0.04) \\ 
  2 & 0.803 (0.71) & 0.197 (0.11) & 0.122 (0.07) & 0.103 (0.05) \\ 
  4 & 2.443 (1.64) & 0.536 (0.40) & 0.251 (0.14) & 0.183 (0.08) \\
  8 & 3.268 (3.60) & 2.030 (2.05) & 0.875 (0.59) & 0.496 (0.35)  \\ \hline
  \multicolumn{5}{c}{MVD of DMWA for $\bm{u}_0 = (1, 2)^T$} \\ \hline
  \backslashbox{$\alpha_1^2$}{$\alpha_2^2$} & 1 & 2 & 3 & 4 \\ 
  \hline
 1 & 2.998 (1.55) & 1.501 (1.56) & 0.699 (1.33) & 0.318 (0.84) \\
 2 & 2.776 (1.42) & 2.028 (1.40) & 1.365 (1.57) & 0.815 (1.18) \\
 4 & 3.213 (1.69) & 2.661 (1.67) & 2.236 (1.67) & 1.644 (1.47) \\
 8 & 3.579 (1.70) & 2.921 (1.39) & 2.700 (1.53) & 2.518 (1.34) \\ \hline
\end{tabular}
\begin{tabular}{ |ccccc| }
 \multicolumn{5}{c}{MVD of STDM for $\bm{u}_0 = (3, 5)^T$}   \\ \hline
  \backslashbox{$\alpha_1^2$}{$\alpha_2^2$} & 1 & 2 & 3 & 4 \\ 
  \hline
  1 & 0.581 (1.16) & 0.403 (1.21) & 0.355 (0.88) & 0.291 (0.99) \\ 
  2 & 1.073 (1.14) & 0.244 (0.13) & 0.142 (0.07) & 0.107 (0.06) \\ 
  4 & 2.781 (2.21) & 0.690 (0.65) & 0.307 (0.18) & 0.211 (0.11) \\
  8 & 3.175 (3.46) & 2.721 (2.55) & 1.123 (0.97) & 0.665 (0.53)  \\ \hline
  \multicolumn{5}{c}{MVD of DMWA for $\bm{u}_0 = (3, 5)^T$} \\ \hline
  \backslashbox{$\alpha_1^2$}{$\alpha_2^2$} & 1 & 2 & 3 & 4 \\ 
  \hline
 1 & 5.100 (2.54) & 2.830 (2.46) & 0.688 (1.54) & 0.098 (0.62) \\
 2 & 5.306 (2.43) & 3.417 (2.51) & 1.819 (2.16) & 1.028 (1.77) \\
 4 & 5.724 (2.47) & 3.951 (2.42) & 3.378 (2.37) & 2.670 (2.32) \\
 8 & 6.109 (2.26) & 5.15 (2.44) & 4.393 (2.33) & 4.061 (2.56) \\ \hline
\end{tabular}%
}
\end{table}

STDM does a better job in estimating moderately small and large wind vectors for the $7 \times 7$ window size compared to DMW Algorithm. Mean vector distance is the smallest when the true spatial range is small and the true temporal range is large. This is intuitive because a small spatial range makes it easier to identify a feature in the target scene, and a large temporal range means that the features dissipate slowly over time. STDM also performs better for the smaller wind vector because when the wind vector is large compared to the window size, the feature tracked in time could potentially move out of the frame, resulting in incorrect wind estimates. In particular, windows of size $7 \times 7$ are not adequate to contain the features being tracked in frame for a wind vector of $(3, 5)^T$ which is reflected in the high MVD and SD values for STDM (see Table \ref{3t1}, top panel).

The performance of the DMW Algorithm also improves as the temporal range increases. However, these simulation results bring forth a major flaw in the DMW Algorithm. The estimated motion winds from the Algorithm are at most half integers and they are limited to the size of the larger search window. That is, while estimating motion vectors, the smaller central target scene ($3 \times 3$) can only move up to 4 pixels in all directions while it is being tracked back and forward in time. As a result, it performs poorly for the larger wind motion vector, which had a v-component of 5 pixel units. This can also be attributed to the window size relative to the magnitude of the wind vector. 

To examine the effect of window size, we now perform the simulations again with window sizes of $11 \times 11$ and $15 \times 15$ respectively, under the same covariance parameter settings as described earlier. The panels of Tables \ref{3t2} and \ref{3t3} portray a clear picture of the effect of window size on the estimation of winds using STDM. The estimation of both wind vectors improve as we increase window size from $7 \times 7$ to $11 \times 11$ (see Table \ref{3t2}) and then to $15 \times 15$ (see Table \ref{3t3}). 

In this part of the simulation study, the true wind has been chosen in a single window. This is analogous to constant wind vectors over a spatial domain. The next two simulation studies consider non-constant wind fields and investigate the effect of the choice of window size and spatial smoothing on wind estimates. 

\subsection{Choosing appropriate window size}

In this part of the simulation study, we examine a simple non-constant wind field to demonstrate that choosing an arbitrarily large target window might lead to inaccurate estimates of the wind vectors. We consider a scenario where the winds diverge from the central meridian of a unit square spatial domain, that is, winds blow west in the west and east in the east. We assume that the wind speed increases quadratically from the center towards the respective edges. Mathematically, we have, 
$$
u_1(\bm{x}, t) = \mbox{sign}(x_1 - 0.5) \times (x_1 - 0.5)^2, \hspace{0.5cm} u_2(\bm{x}, t) = 0.
$$

Based on this wind pattern, data sets are generated on a $[0, 1]^2$ spatiotemporal grid of size $100 \times 100 \times 3$ using the aforementioned fast Gaussian approximation algorithm with an exponential space-time covariance function ($M_{\nu}$ with $\nu = 1/2$, $\alpha_1^2 = 0.075$ (4 pixels) and $\alpha_2^2 = 5$). We fit the proposed space-time drift model locally in space and estimate all parameters using multiple window sizes. Since we assume that the wind field does not vary along the vertical axis, the estimation is done along the horizontal axis at a subset of locations along the vertical axis. 

\begin{figure}  % spans both columns
\centering
\begin{minipage}{0.47\linewidth}
    \centering
    \includegraphics[width=\linewidth]{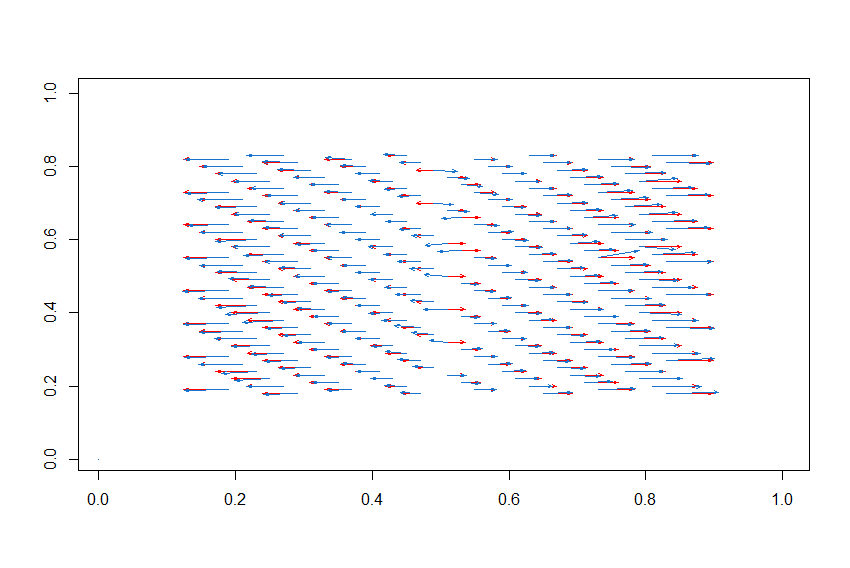}
    \subcaption{Wind estimates from STDM ($25 \times 25$)}
\end{minipage}
\begin{minipage}{0.47\linewidth}
\includegraphics[width=\linewidth]{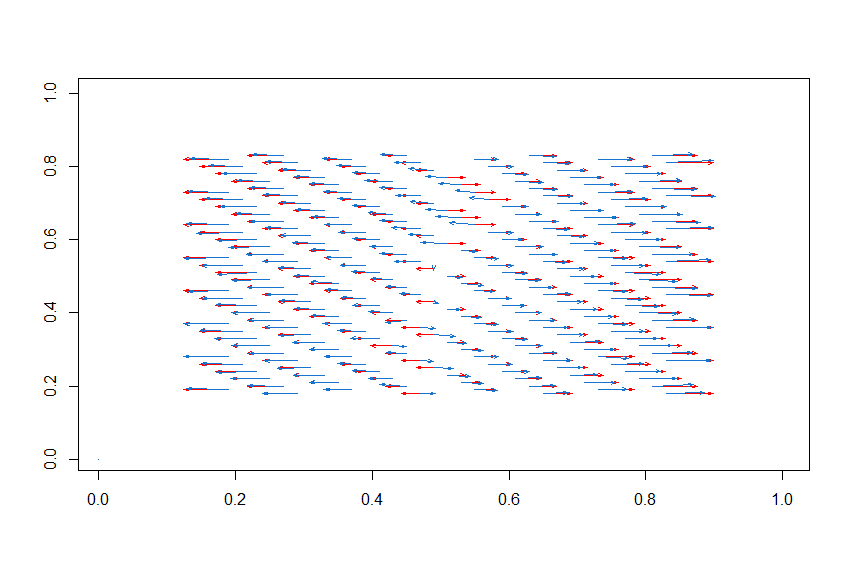}
\subcaption{Wind estimates from STDM ($35 \times 35$)}
\end{minipage}
\begin{minipage}{0.47\linewidth}
\includegraphics[width=\linewidth]{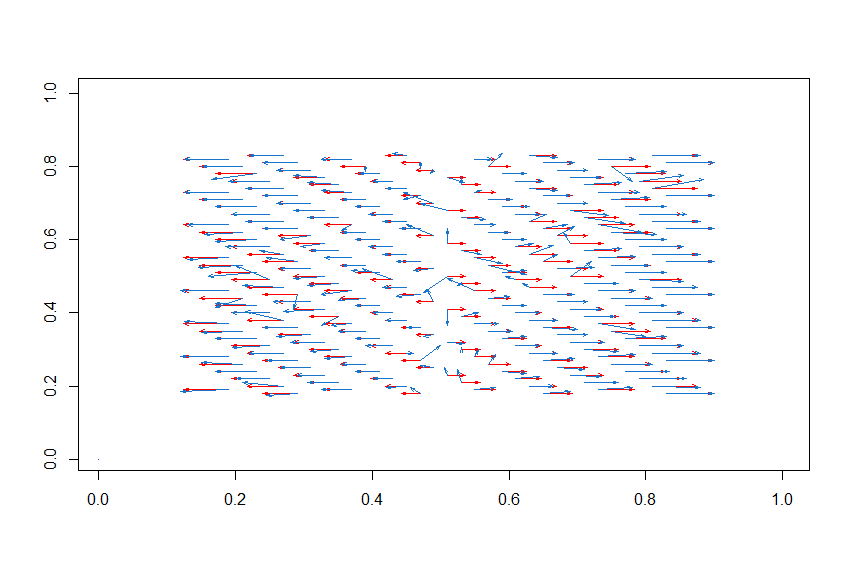}
\subcaption{Wind estimates from DMWA ($25 \times 25$)}
\end{minipage}
\begin{minipage}{0.47\linewidth}
\includegraphics[width=\linewidth]{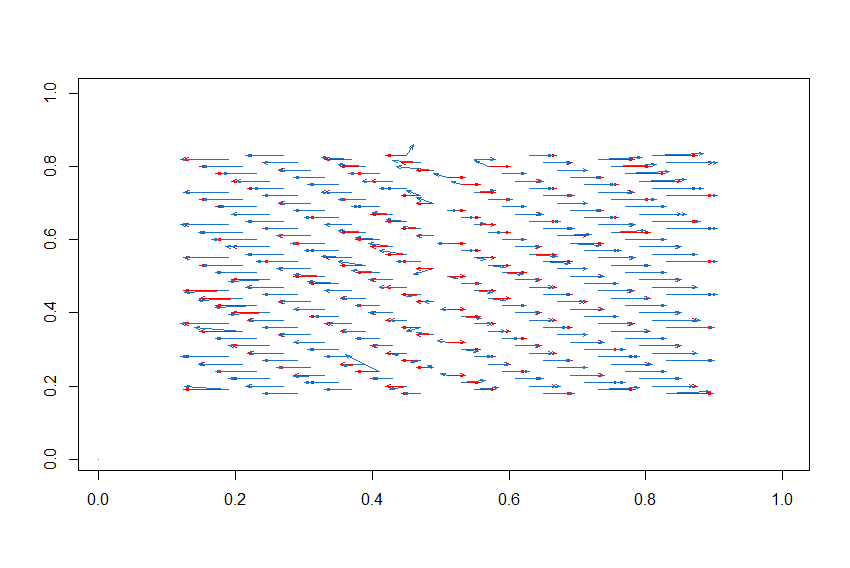}
\subcaption{Wind estimates from DMWA ($35 \times 35$)}
\end{minipage}
\caption{True (red) and estimated (blue) wind fields using the Space-time Drift Model (top row) and the DMW Algorithm (bottom row) for one representative data set, simulated using non-constant wind field. The left column shows wind estimates calculated using window of size $25 \times 25$ and the right column shows the same calculated using window of size $35 \times 35$. The wind vectors are plotted for a subset of spatial locations for clear visualization.} % Overall figure caption
\label{fig1}
\end{figure}

Table \ref{mytable2} compares the average performance of the two methods over $N = 50$ simulated data sets. From the results, it can be seen that while STDM performs better than the DMW Algorithm for all window sizes, the performance of STDM deteriorates for windows of sizes larger than $25 \times 25$. The estimated wind fields from both methods are shown in Figure \ref{fig1}. From Figure \ref{fig1}, it can be seen that the estimates of the smaller wind vectors on both sides of the central meridian become less accurate with increasing window size. This accounts for the increase of the MVD and the associated SD for window sizes larger than $25 \times 25$.

\begin{table}[bt]
\vspace*{1 cm}
\caption{\label{mytable2}Comparing Mean Vector Difference (SD) for wind estimates from the Space-Time Drift Model (STDM) and the Derived Motion Winds (DMW) Algorithm averaged over $N = 50$ data sets simulated using non-constant wind field. MVD and SD are measured in pixels.}
\centering
\resizebox{\textwidth}{!}{
\begin{tabular}{ cccccccc }
Method & $7 \times 7$ & $11 \times 11$ & $15 \times 15$ & $21 \times 21$ & $25 \times 25$ & $31 \times 31$ & $35 \times 35$\\ \hline
STDM &   3.128 (3.588) &  1.537 (2.271) & 0.905 (1.413) & 0.661 (1.153) & 0.619 (1.093) & 0.707 (1.139) & 0.818 (1.280)  \\
DMWA & 4.140 (1.681) & 3.414 (2.011) & 2.926 (2.180) & 1.935 (2.007) & 1.214 (1.517) & 0.998 (1.446) & 0.830 (1.320) \\ \hline
\end{tabular}}
\end{table}

\subsection{Efficacy of spatial smoothing} We now conduct another simulation study which examines the performance of the competing methods and looks into the efficacy of spatial smoothing of wind estimates. Once again, we generate data sets on a $100 \times 100$ unit square at three consecutive time points. While generating the data, we consider rotational winds, which are obtained by rotating each location counter-clockwise by 6$\degree$ around the origin at each time point. If $\bm{x}$ denotes a spatial location at time $t$ and $\bm{x}'$ denotes the same at time $(t + 1)$, then, 
$$
\begin{pmatrix} x_1' \\ x_2' \end{pmatrix} = \begin{pmatrix} \mbox{cos}\theta & -\mbox{sin}\theta \\ \mbox{sin}\theta & \mbox{cos}\theta\end{pmatrix} \begin{pmatrix} x_1 \\ x_2 \end{pmatrix}, 
$$
for $t = 1, 2$ and $\theta = 6 \pi/180$. Then 
$$
u_1(\bm{x}, t) = x_1' - x_1 \hspace{0.25cm} \text{and} \hspace{0.25cm}  u_2(\bm{x}, t) = x_2' - x_2.
$$
These winds are non-uniform over space and emulate big tropical storms which rotate counter-clockwise in the Northern Hemisphere and clockwise in the Southern Hemisphere due to the Coriolis effect. This simulation setting reflects realistic scenarios and compensate for the lack of availability of data sets with ground truth for wind measurements.

Once the shifted locations are obtained at three consecutive time points, we use the `GpGp' package \citep{guinness2018gpgp} in R to generate the data sets for the three consecutive time points. The covariance function used for generating the data is the exponential space-time covariance with true spatial range $\alpha_1^2 = 0.075$ (4 pixels) and true temporal range $\alpha_2^2 = 5$. We estimate the parameters locally using the STDM at $60 \times 60$ locations at the center of the grid. This reduces boundary effects during local estimation and allows us to compare the performance of our method for multiple window sizes. The wind estimates are spatially smoothed using a weighted Gaussian kernel. We also estimate the winds using DMW Algorithm under similar settings, and the DMW estimates are also smoothed for fair comparison. 

Table \ref{mytable3} compares the average performance of the algorithms and their smoothed versions over $N = 50$ simulated data sets. From the results, it can be seen that the STDM outperforms the DMW Algorithm for all window sizes. Also, smoothing the estimates gives us a much better representation of the wind fields compared to the raw estimates.

\begin{figure}  % spans both columns
\centering
\begin{minipage}{0.48\linewidth}
\includegraphics[width=\linewidth]{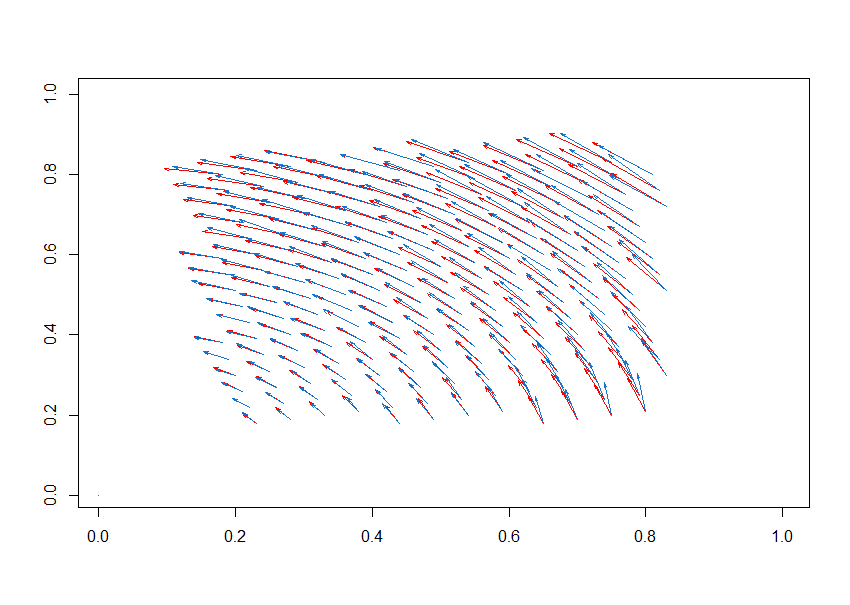}
\subcaption{Raw wind estimates from STDM}
\end{minipage}
\begin{minipage}{0.48\linewidth}
\includegraphics[width=\linewidth]{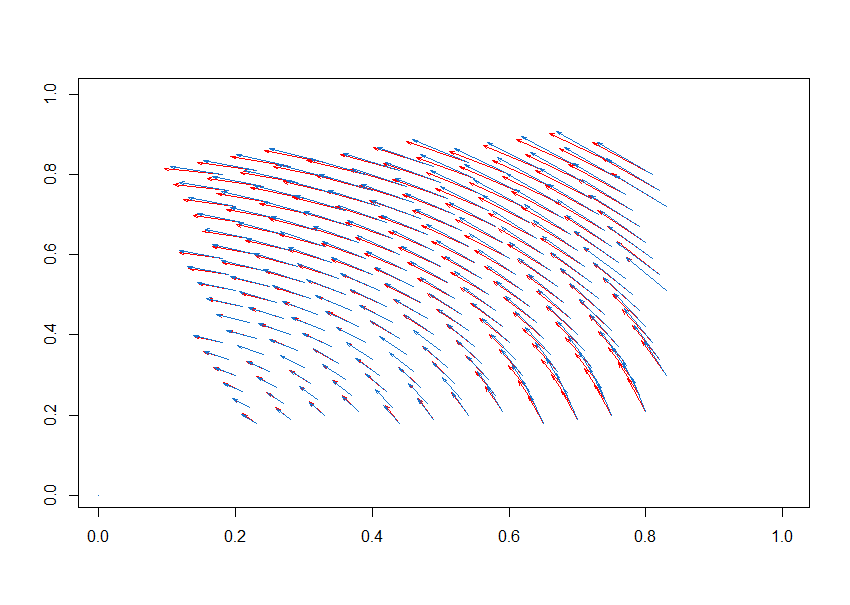}
\subcaption{Smoothed wind estimates from STDM}
\end{minipage}
\begin{minipage}{0.48\linewidth}
\includegraphics[width=\linewidth]{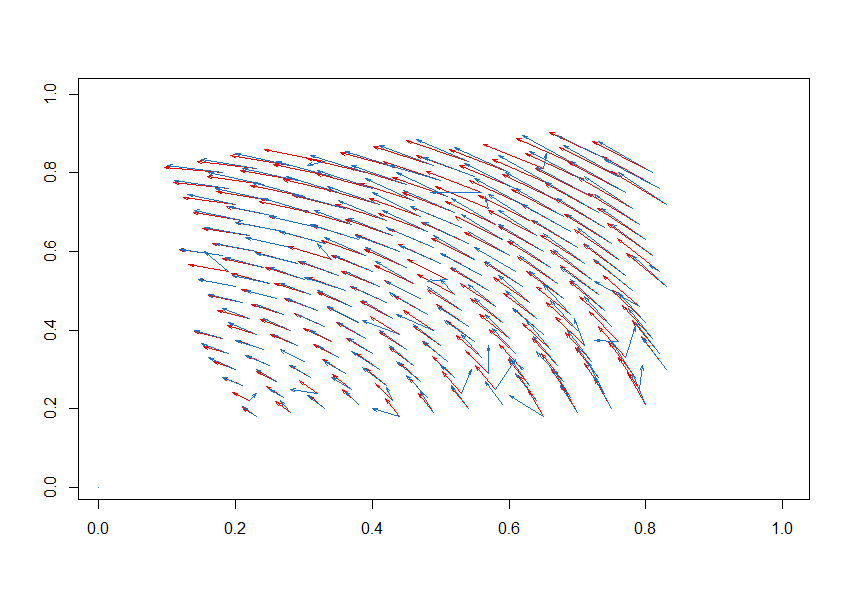}
\subcaption{Raw wind estimates from DMWA}
\end{minipage}
\begin{minipage}{0.48\linewidth}
\includegraphics[width=\linewidth]{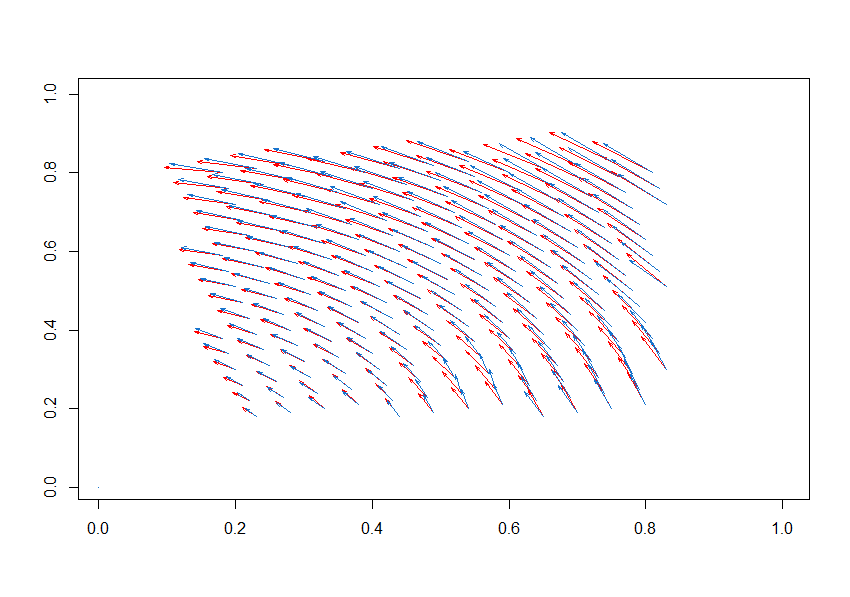}
\subcaption{Smoothed wind estimates from DMWA}
\end{minipage}
\caption{True (red) and estimated (blue) wind fields using the Space-time Drift Model (top row) and the DMW Algorithm (bottom row) for one representative data set, simulated using rotational wind field. The left column shows raw wind estimates calculated using window size $25 \times 25$ and the right column shows the corresponding smoothed estimates. The wind vectors are plotted for a subset of spatial locations for clear visualization.} % Overall figure caption
\label{fig2}
\end{figure}

As discussed in Section 4.2, the size of the target window plays a crucial role while estimating motion winds. For example, while estimating the rotational wind field using $15 \times 15$ (or smaller) windows, some of the target features move out of scene because of the non-uniformity in winds across space. In that scenario, both methods provide poor estimates of the wind field. Using larger target windows, such as windows of size $25 \times 25$, fixes the problem to some extent. For rotational wind field, larger window size improves estimation of the wind fields. This is because the average of the rotational wind field over a window is equal to the wind vector at the center of the window. Figure \ref{fig2} shows the respective true wind fields (in red) along with the raw and smoothed estimates (in blue) from both methods for a single representative data set obtained using windows of size $25 \times 25$. The wind vectors are plotted for a subset of the $60^2$ locations in the spatial domain for clear visualization of the true and estimated fields. 

A major takeaway from the simulation studies is that while estimating non-uniform wind fields, selecting the window size requires balancing a trade-off between a window size that is large enough to capture targets moving through the scene yet small enough to satisfy the assumption that within the window the process is stationary with a constant drift. This motivates us to choose the window size using cross validation in the real-data analysis.

\begin{table}[bt]
\vspace*{1 cm}
\caption{\label{mytable3}Comparing Mean Vector Difference (SD) for raw and smoothed wind estimates from the Space-Time Drift Model (STDM) and the Derived Motion Winds (DMW) Algorithm averaged over $N = 50$ data sets simulated using rotational wind field. MVD and SD are measured in pixels.}
\centering
\resizebox{\textwidth}{!}{
\begin{tabular}{ cccccc }
Method & $7 \times 7$ & $11 \times 11$ & $15 \times 15$ & $21 \times 21$ & $25 \times 25$ \\ \hline
STDM &   7.375 (6.007) &  5.357 (5.450) &  3.903 (4.745) & 2.414 (3.373) &  1.058 (1.854) \\
DMWA & 7.551 (2.418) & 7.065 (3.056) & 6.143 (3.778) & 3.267 (3.440) & 1.371 (1.937) \\
Smoothed STDM &  6.158 (3.602) & 4.337 (3.925) & 2.659 (3.179) & 1.432 (1.733) & 0.571 (0.528) \\
Smoothed DMWA &  7.473 (2.156) & 6.912 (2.616) & 5.776 (2.958) & 2.711 (2.095)  & 0.959 (0.735) \\ \hline
\end{tabular}}
\end{table}

\section{Application to GOES-15 data}

\subsection{GOES-15 data description}
The Geostationary Operational Environmental Satellite (GOES), operated by National Oceanic and Atmospheric Administration (NOAA) provides continual measurements of the atmosphere and surface variables, which help facilitate meteorological research including weather forecasting and severe storm tracking. Since the launch of GOES-8 in 1994, the GOES instruments have monitored atmospheric phenomena and provided a continuous stream of environmental data. The data set used in this project is from the GOES-15 satellite. Launched in March 2010, GOES-15 is positioned at the GOES-West location of 135\degree W longitudes over the Pacific Ocean. The data set, as described in \citet{knapp2018gridded}, is a gridded satellite Contiguous US domain data which are geostationary data remapped to equal angle projection with an 0.04\degree ($\sim$4 km) latitudinal resolution and 15 minutes temporal resolution. The data set includes infrared channel data in terms of pixel-wise brightness temperature for the reflective bands (channels 1 - 6 with approximate central wavelengths 0.47, 0.64, 0.865, 1.378, 1.61, 2.25 microns respectively). The reflective bands support among other ground and atmospheric indicators, the characterization of clouds. Gridded GOES-15 data can be obtained from NOAA One-Stop at \href{https://data.noaa.gov/onestop/#/collections/details/AWbwYbuZHtVRBOAZkAzC?q=GOES}{https://data.noaa.gov/onestop/\#/collections}.

\begin{figure}
\centering
\begin{minipage}{0.32\textwidth}
\includegraphics[width=\linewidth]{Rplot02t1}
\end{minipage} %
\begin{minipage}{0.32\linewidth}
\includegraphics[width=\linewidth]{Rplot02t2}
\end{minipage} %
\begin{minipage}{0.345\linewidth}
\includegraphics[width=\linewidth]{Rplot02t3}
\end{minipage}%

\begin{minipage}{0.32\textwidth}
\includegraphics[width=\linewidth]{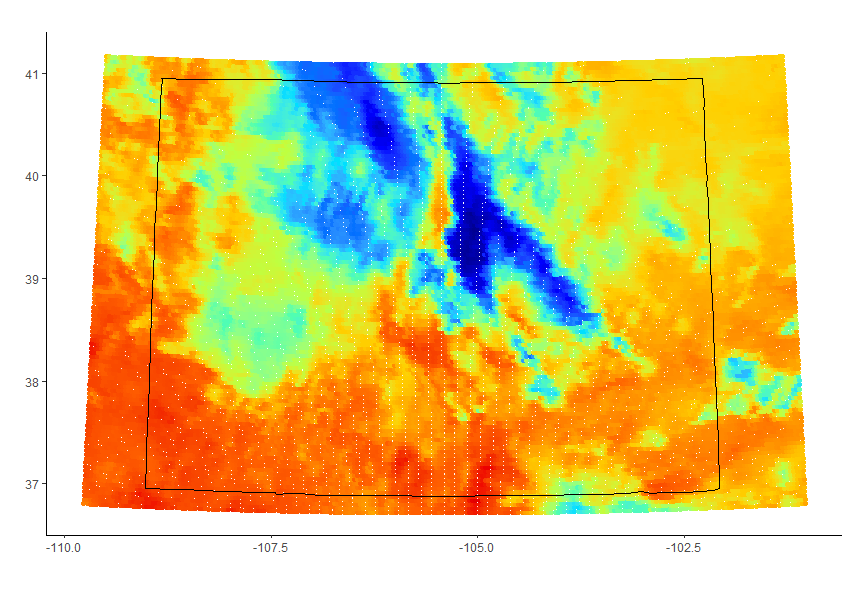}
\end{minipage} %
\begin{minipage}{0.32\linewidth}
\includegraphics[width=\linewidth]{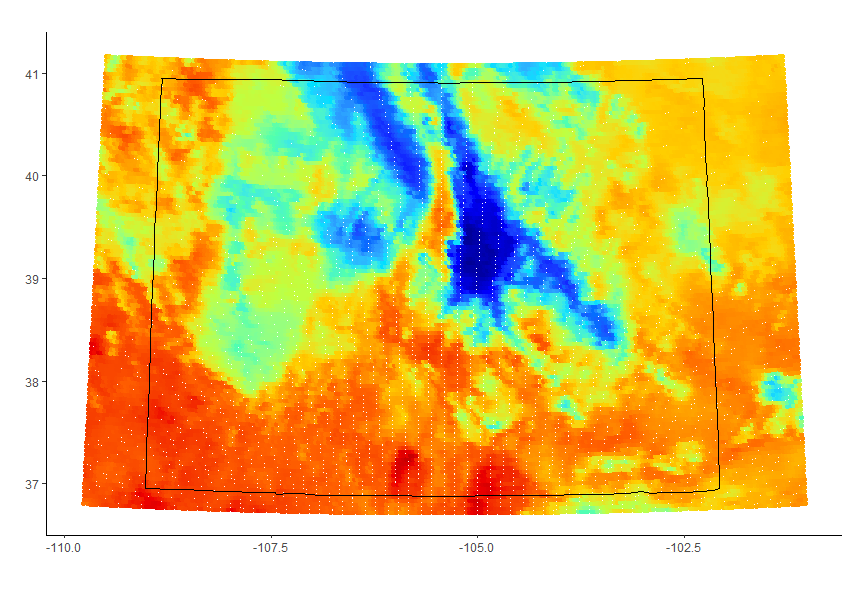}
\end{minipage} %
\begin{minipage}{0.345\linewidth}
\includegraphics[width=\linewidth]{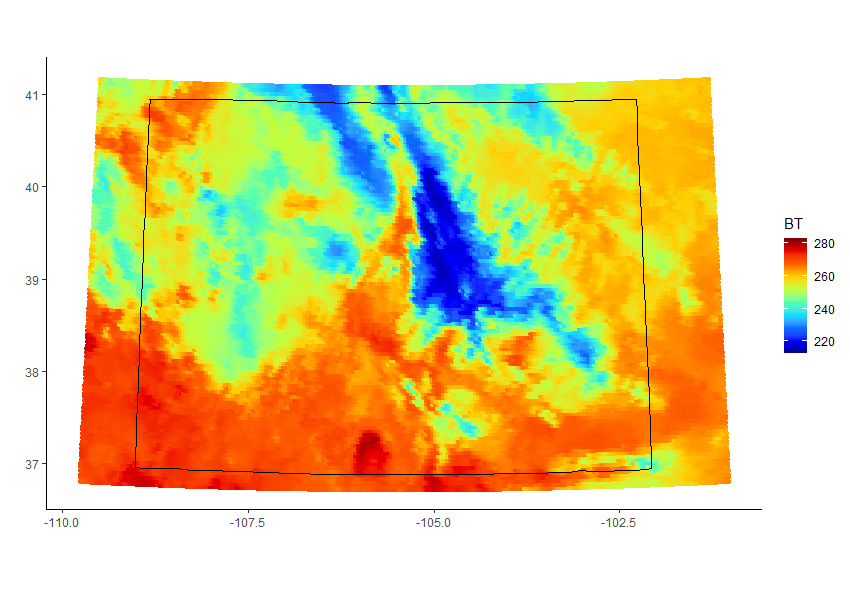}
\end{minipage}
\caption{ Data buffers containing brightness temperature (Kelvin) maps over Colorado on January 3, 2015 at 00:15 am (top panel) and on January 4, 2015 at 08:45 am (bottom panel).}
\label{fig2a}
\end{figure}

We analyze Channel 4 brightness temperature data (recorded in Kelvin scale) for 10 consecutive days starting January 1, 2015 at a temporal resolution of 15 minutes (960 total images), covering the state of Colorado ($36.82 \degree$ N to $41.18 \degree$ N latitudes and $109.78 \degree$ W to $101.02 \degree$ W longitudes). Figure \ref{fig2a} shows the data in two representative data buffers centered at 00:15 am on January 03, 2015 (t = 193; top panel) and at 08:45 am on January 04, 2015 (t = 322; bottom panel) respectively. These two time points represent wind movements across the state at night and in the morning respectively. Lower values of brightness temperature indicate cloud cover, whereas high brightness temperature values suggest clear skies over the region. 

\subsection{Estimation using GOES-15 data} 

The first step involves standardizing the brightness temperature data using the pixel-wise sample mean and standard deviation over time. This removes the constant background image and the effect of low cloud cover over the region which can affect the local estimation of wind vectors. To smooth the standard deviation map, we use a Gaussian kernel with smoothing parameter $\lambda = 3$ pixels (i.e., 12 km).  

\begin{figure}  % spans both columns
    \centering
    \begin{minipage}{0.48\linewidth}
        \centering
        \includegraphics[width=\linewidth]{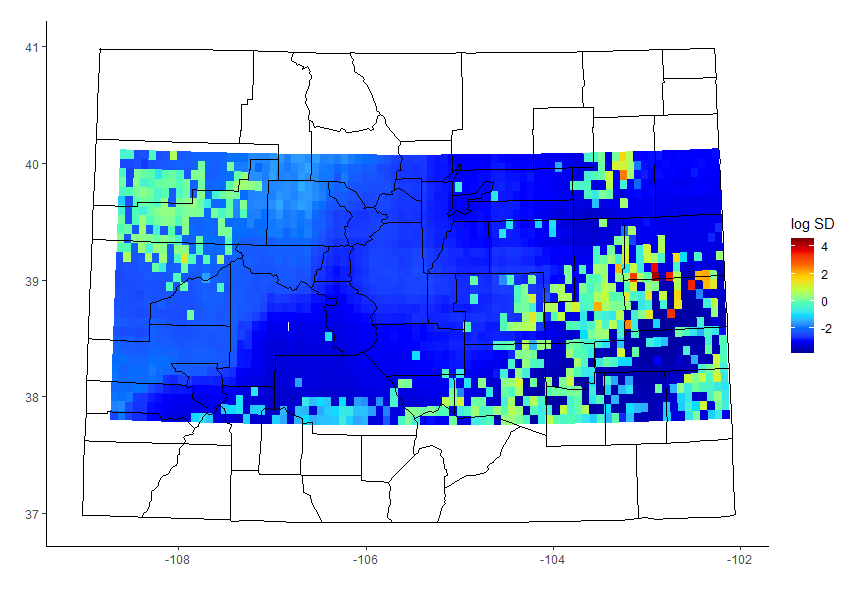}
        \subcaption{Estimated log SD for $\widehat{u}_1$ (Jan 3; 00:15 am)}
    \end{minipage}
    \begin{minipage}{0.48\linewidth}
        \centering
        \includegraphics[width=\linewidth]{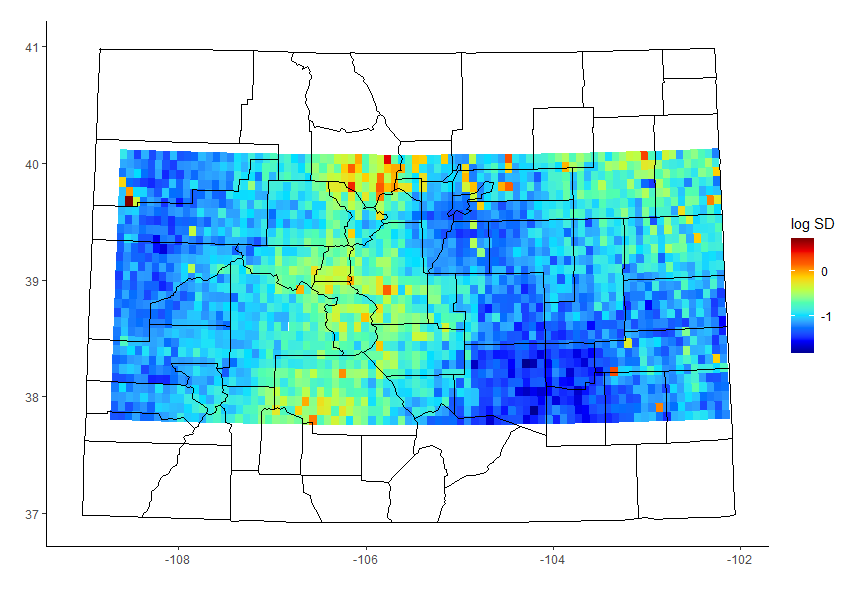}
        \subcaption{Estimated log SD for $\widehat{u}_1$ (Jan 4; 08:45 am)}
    \end{minipage}
    \begin{minipage}{0.48\linewidth}
        \centering
        \includegraphics[width=\linewidth]{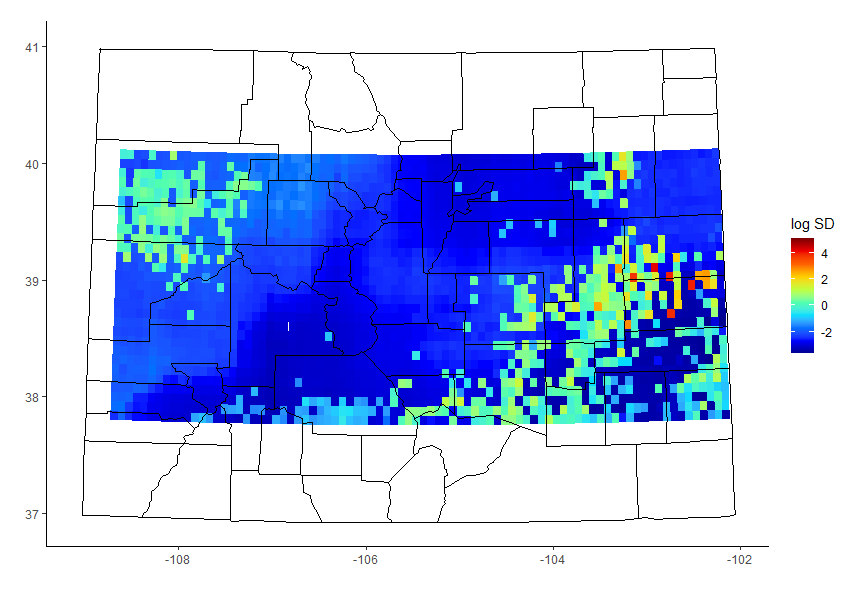}
        \subcaption{Estimated log SD for $\widehat{u}_2$ (Jan 3; 00:15 am)}
    \end{minipage}
    \begin{minipage}{0.48\linewidth}
        \centering
        \includegraphics[width=\linewidth]{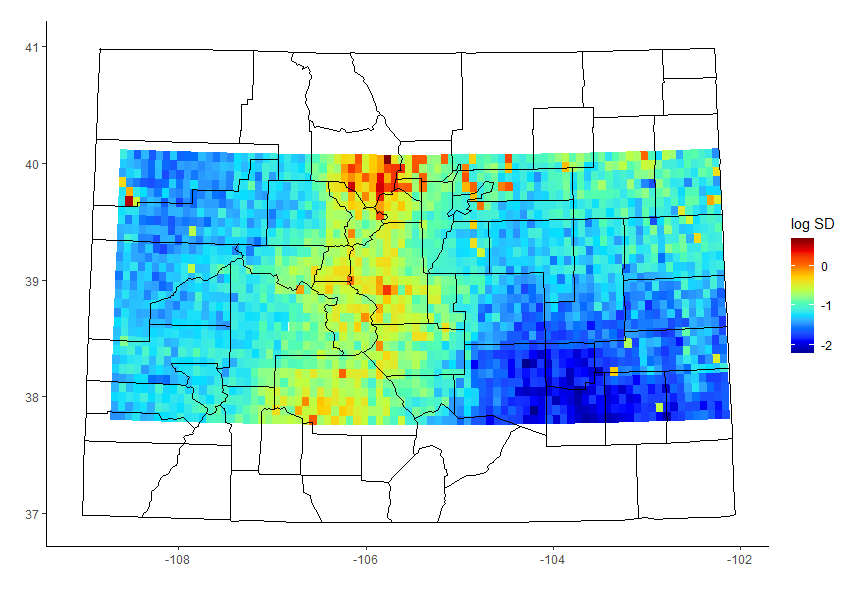}
        \subcaption{Estimated log SD for $\widehat{u}_2$ (Jan 4; 08:45 am)}
    \end{minipage}
    \caption{Estimated standard deviations (in log scale) obtained using the Space-Time Drift Model (STDM), corresponding to estimated east-west (top row) and north-south (bottom row) wind components on January 3, 2015 at 00:15 am (left column) and January 4, 2015 at 08:45 am (right column) respectively.} % Overall figure caption
    \label{fig3.3}
\end{figure}

We fit a Gaussian process with the full covariance function in \eqref{eq3.2} to the standardized data within each target window using the `matern\_anisotropic3D\_alt' covariance function in the `GpGp' package in R. As mentioned earlier, the covariance function in \eqref{eq3.2} includes an additional parameter $a_{12}$ to allow for anisotropy in the spatial covariances. We use cross validation to determine the appropriate size of target windows. Since no direct measurements of winds are available, we compare window sizes indirectly based on predictions of brightness temperature at the fourth time point using the wind estimate based on the first three time points and use this optimal window size to estimate winds for all subsequent time points. Table \ref{tableCV} gives the Mean Squared Prediction Error (MSPE) (discussed in Section 5.3) for different window sizes and the corresponding average computation time (in seconds) to estimate the wind vector at one pixels using 3 consecutive time steps. All computations are done on a PC with Intel Core i7-9750H CPU at 2.60GHz with 32 GB of RAM. Based on the results, we select sub regions of size $25 \times 25$ pixels (i.e., 100 km $\times$ 100 km) and estimate the wind vectors, along with all other covariance parameters, locally at each spatial location and at each time using maximum likelihood estimates. We also estimate the variances associated with the estimates by computing the Fisher information at the MLE. Figure \ref{fig3.3} shows the uncertainty associated with the estimated wind components as given by the estimated standard deviations in the log scale, for the two data buffers mentioned above.

\begin{table}[bt]
\caption{\label{tableCV}Mean Squared Prediction Error (MSPE) for cross validation based on the first three time points and the corresponding average computation time (in seconds) per pixel}
\centering
    \begin{tabular}{ccc} 
      window size  & MSPE & Comp. time (in secs) \\ \hline 
       $11 \times 11$ & 0.398  & 13.38\\   
       $15 \times 15$ & 0.370  & 27.90 \\
       $21 \times 21$ & 0.384  & 54.21 \\
       $25 \times 25$ & 0.344  & 72.41 \\
       $35 \times 35$ & 0.373  & 94.28 \\ \hline
    \end{tabular}
\end{table}

%\begin{figure}  % spans both columns
%\centering
%\begin{minipage}{0.315\linewidth}
%\includegraphics[width=\linewidth]{Rplotsdu2}
%\end{minipage}
%\begin{minipage}{0.315\linewidth}
%\includegraphics[width=\linewidth]{Rplotsdu3}
%\end{minipage}
%\begin{minipage}{0.33\linewidth}
%\includegraphics[width=\linewidth]{Rplotsdu4}
%\end{minipage}

%\begin{minipage}{0.315\linewidth}
%\includegraphics[width=\linewidth]{Rplotsdv2}
%\end{minipage}
%\begin{minipage}{0.315\linewidth}
%\includegraphics[width=\linewidth]{Rplotsdv3}
%\end{minipage}
%\begin{minipage}{0.33\linewidth}
%\includegraphics[width=\linewidth]{Rplotsdv4}
%\end{minipage}
%\caption{Estimated standard deviations (in log scale) obtained using the Space-Time Drift Model (STDM), corresponding to estimated east-west (top row) and north-south (bottom row) wind components at three consecutive time points (represented by the columns from left to right).} % Overall figure caption
%\label{fig3.3}
%\end{figure}

The estimated variances are high at some locations, especially where it is difficult to distinguish clouds from the background terrain and track their movements across the region. To account for this, we smooth each component of the estimated wind field using weighted Gaussian kernels, the weights being scaled to the inverse variances of the estimates (details are given in the Appendix). The smoothing parameter has been chosen based on cross validation. 

\begin{figure}  % spans both columns
    \centering
    \begin{minipage}{0.48\linewidth}
        \centering
        \includegraphics[width=\linewidth]{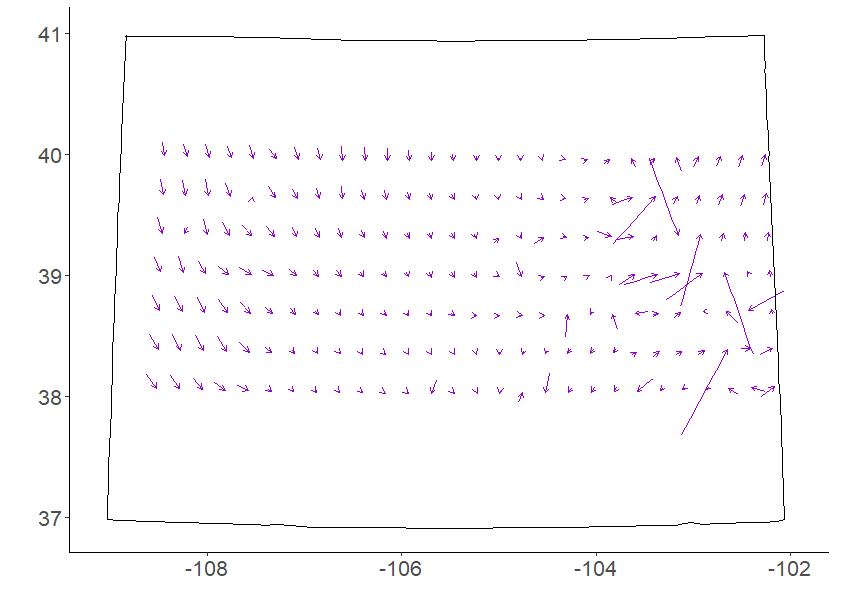}
        \subcaption{Initial wind estimates (Jan 3; 00:15 am)}
    \end{minipage}
    \begin{minipage}{0.48\linewidth}
        \centering
        \includegraphics[width=\linewidth]{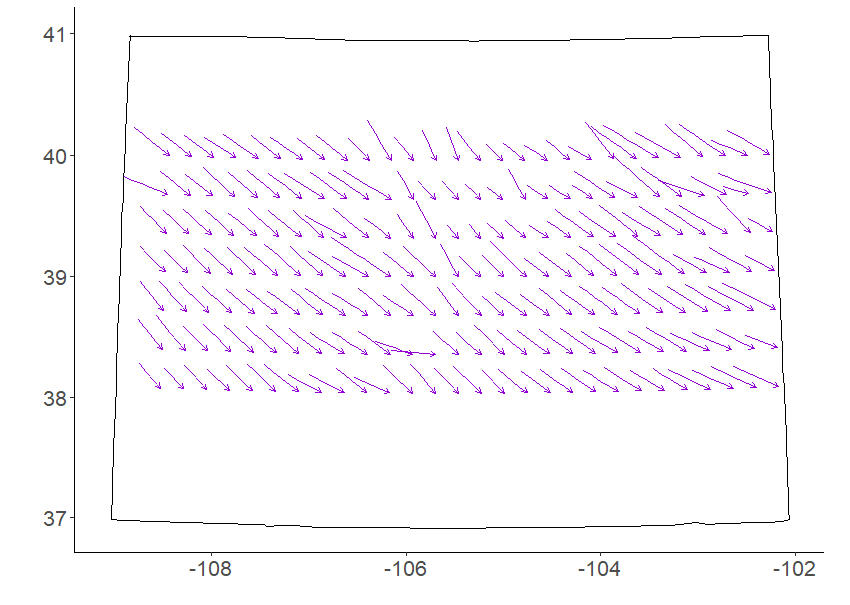}
        \subcaption{Initial wind estimates (Jan 4; 08:45 am)}
    \end{minipage}
    \begin{minipage}{0.48\linewidth}
        \centering
        \includegraphics[width=\linewidth]{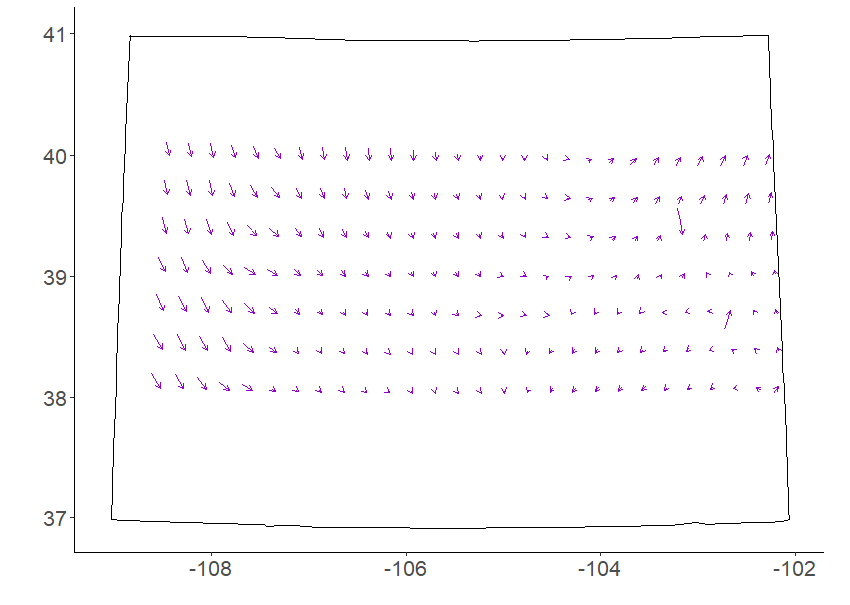}
        \subcaption{Smoothed wind estimates (Jan 3; 00:15 am)}
    \end{minipage}
    \begin{minipage}{0.48\linewidth}
        \centering
        \includegraphics[width=\linewidth]{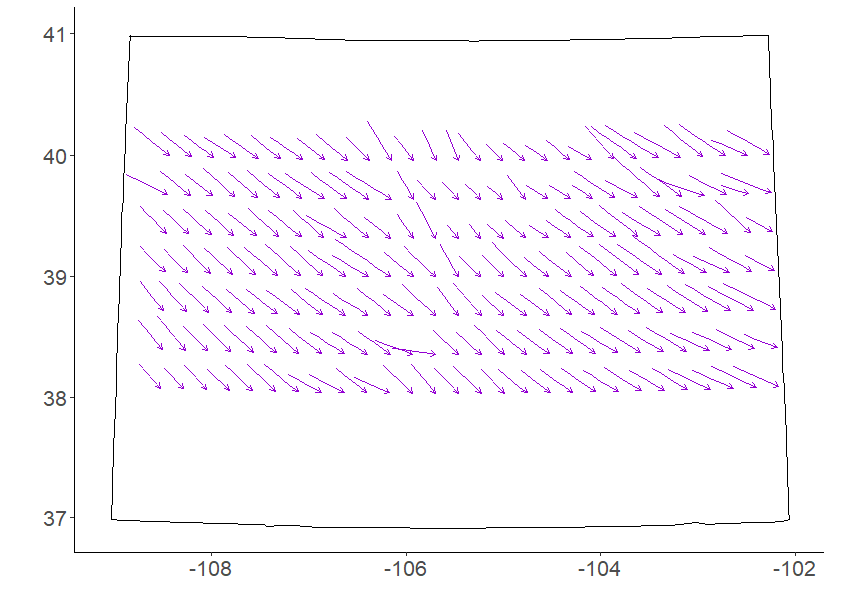}
        \subcaption{Smoothed wind estimates (Jan 4; 08:45 am)}
    \end{minipage}
    \caption{Initial (top row) and smoothed (bottom row) wind field estimates (km/15 mins) over Colorado obtained using the Space-Time Drift Model (STDM). The left column shows estimated wind on January 3, 2015 at 00:15 am and the right column shows the same on January 4, 2015 at 08:45 am. The wind vectors are plotted for a subset of spatial locations for clear visualization.} % Overall figure caption
    \label{fig3.4}
\end{figure}

Similar to the DMW Algorithm, we also impose some simple criteria to check whether a target window $D(\bm{x}, t)$ is suitable for tracking and estimating motion winds. We look at the mean of the residual variances from regressing $Z_D(\cdot, t)$ on $Z_D(\cdot, t - 1)$ and $Z_D(\cdot, t + 1)$ on $Z_D(\cdot, t)$. We say that $D(\bm{x}, t)$ is discarded if the mean residual variance is low and the brightness temperature at the location, $Y(\bm{x}, t)$ is high. This signifies no cloud cover or negligible movement across time within the target window. For our data, we reject $D(\bm{x}, t)$ if the mean residual variance is in the lower 10 percentile of all mean residual variances and the brightness temperature $Y(\bm{x}, t)$ is in its upper 10 percentile, which is, $Y(\bm{x}, t) > 260$ Kelvin. Figure \ref{fig3.4} shows the raw and smoothed wind estimates obtained from the proposed STDM. The wind vectors have been plotted at a subset of spatial locations for better visualization. 

\begin{figure}  % spans both columns
    \centering
    \begin{minipage}{0.48\linewidth}
        \centering
        \includegraphics[width=\linewidth]{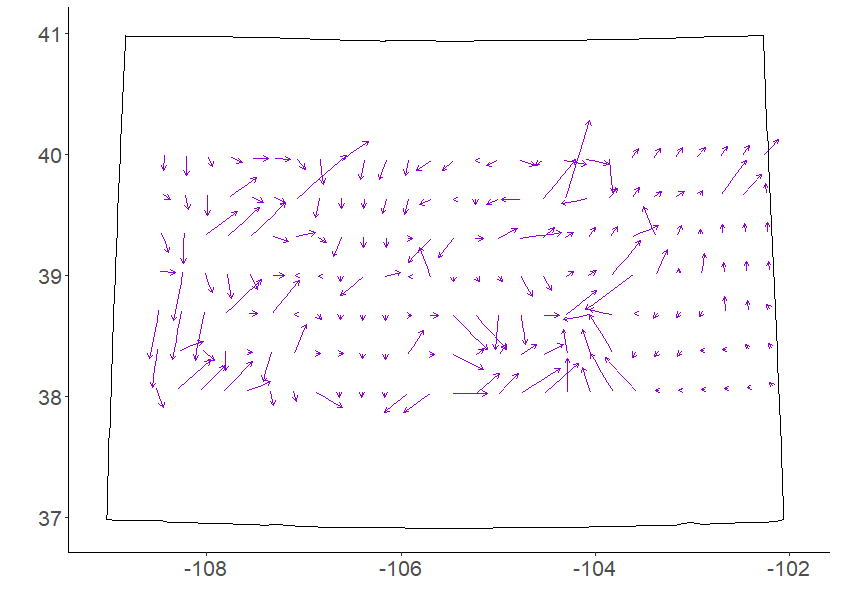}
        \subcaption{Initial wind estimates (Jan 3; 00:15 am)}
    \end{minipage}
    \begin{minipage}{0.48\linewidth}
        \centering
        \includegraphics[width=\linewidth]{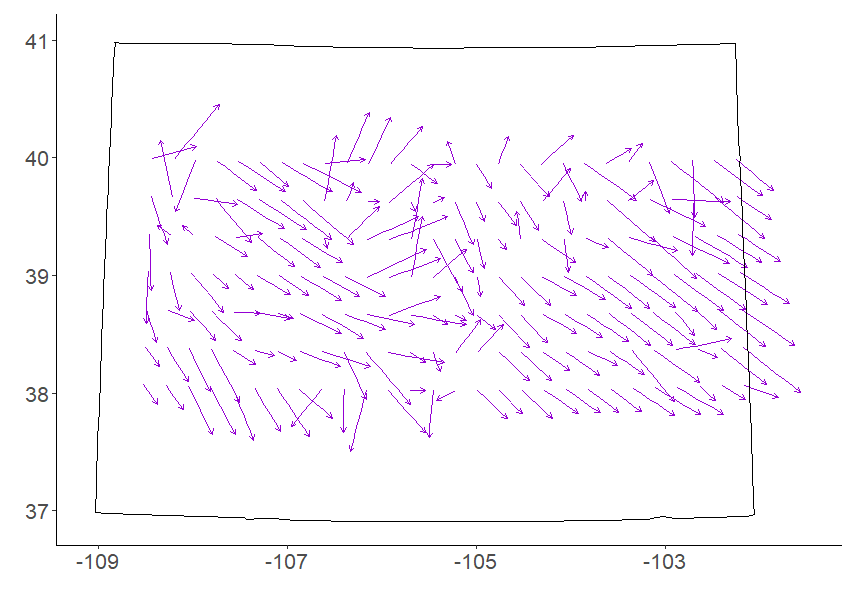}
        \subcaption{Initial wind estimates (Jan 4; 08:45 am)}
    \end{minipage}
    \begin{minipage}{0.48\linewidth}
        \centering
        \includegraphics[width=\linewidth]{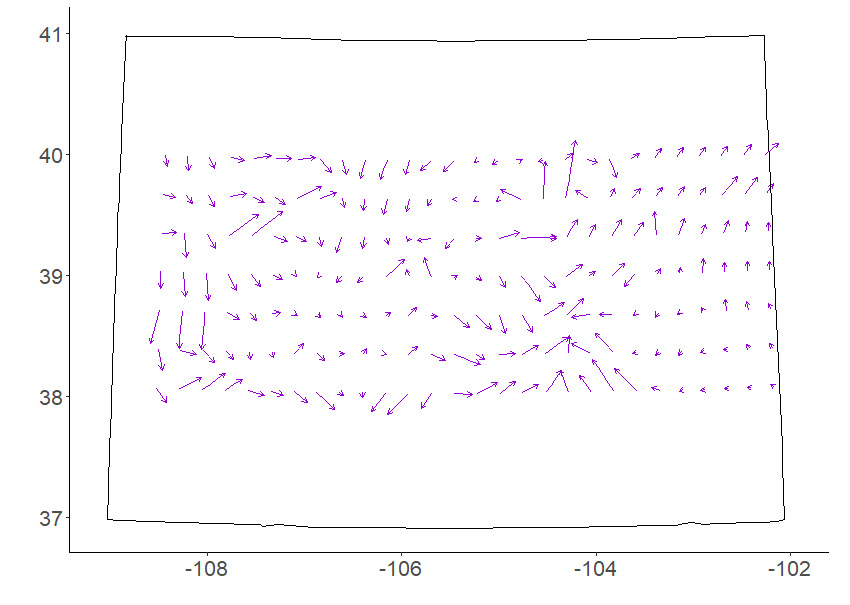}
        \subcaption{Smoothed wind estimates (Jan 3; 00:15 am)}
    \end{minipage}
    \begin{minipage}{0.48\linewidth}
        \centering
        \includegraphics[width=\linewidth]{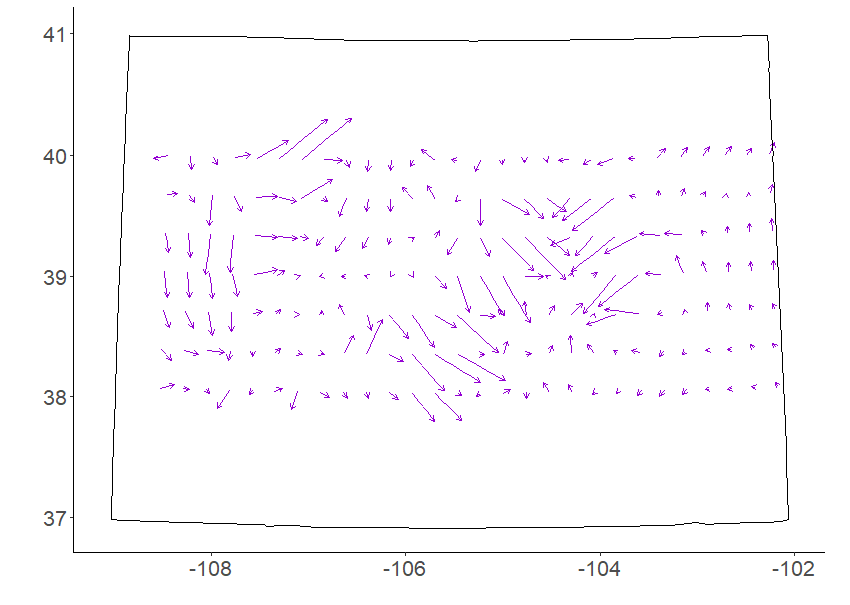}
        \subcaption{Smoothed wind estimates (Jan 4; 08:45 am)}
    \end{minipage}
    \caption{Initial (top row) and smoothed (bottom row) wind field estimates (km/ 15 mins) over Colorado obtained from the DMW Algorithm (DMWA). The left column shows estimated wind on January 3, 2015 at 00:15 am and the right column shows the same on January 4, 2015 at 08:45 am. The wind vectors are plotted for a subset of spatial locations for clear visualization.} % Overall figure caption
    \label{fig3.5}
\end{figure}

For comparison, the DMW estimates are also smoothed using a simple Gaussian kernel smoother with its optimal smoothing parameter chosen using cross validation. The raw and smoothed estimates of the wind fields obtained from DMW Algorithm are plotted at a subset of locations in Figure \ref{fig3.5}.

% \begin{figure}  % spans both columns
% \centering
% \begin{minipage}{0.32\linewidth}
% \includegraphics[width=\linewidth]{Rplotraw2}
% \end{minipage}
% \begin{minipage}{0.32\linewidth}
% \includegraphics[width=\linewidth]{Rplotraw3}
% \end{minipage}
% \begin{minipage}{0.32\linewidth}
% \includegraphics[width=\linewidth]{Rplotraw4}
% \end{minipage}

% \begin{minipage}{0.32\linewidth}
% \includegraphics[width=\linewidth]{Rplotsmth2}
% \end{minipage}
% \begin{minipage}{0.32\linewidth}
% \includegraphics[width=\linewidth]{Rplotsmth3}
% \end{minipage}
% \begin{minipage}{0.32\linewidth}
% \includegraphics[width=\linewidth]{Rplotsmth4}
% \end{minipage}
% \caption{Raw (top row) and smoothed (bottom row) wind field estimates at three consecutive time points obtained using the Space-Time Drift Model (STDM). The columns (from left to right) represent 3 consecutive time points. The wind vectors are plotted for a subset of spatial locations for clear visualization.} % Overall figure caption
% \label{fig3.4}
% \end{figure}

% \begin{figure}  % spans both columns
% \centering
% \begin{minipage}{0.32\linewidth}
% \includegraphics[width=\linewidth]{RplotrawNT2}
% \end{minipage}
% \begin{minipage}{0.32\linewidth}
% \includegraphics[width=\linewidth]{RplotrawNT3}
% \end{minipage}
% \begin{minipage}{0.32\linewidth}
% \includegraphics[width=\linewidth]{RplotrawNT4}
% \end{minipage}

% \begin{minipage}{0.32\linewidth}
% \includegraphics[width=\linewidth]{Rplot02smthNT2}
% \end{minipage}
% \begin{minipage}{0.32\linewidth}
% \includegraphics[width=\linewidth]{Rplot02smthNT3}
% \end{minipage}
% \begin{minipage}{0.32\linewidth}
% \includegraphics[width=\linewidth]{Rplot02smthNT4}
% \end{minipage}
% \caption{Raw (top row) and smoothed (bottom row) wind field estimates at three consecutive time points, obtained from the Derived Motion Winds Algorithm (DMWA). The columns (from left to right) represent 3 consecutive time points. The wind vectors are plotted for a subset of spatial locations for clear visualization.} % Overall figure caption
% \label{fig3.5}
% \end{figure}

\subsection{Comparison based on Mean Squared Prediction Error}

\begin{figure}  % spans both columns
\centering
\begin{minipage}{0.465\textwidth}
\includegraphics[width=\linewidth]{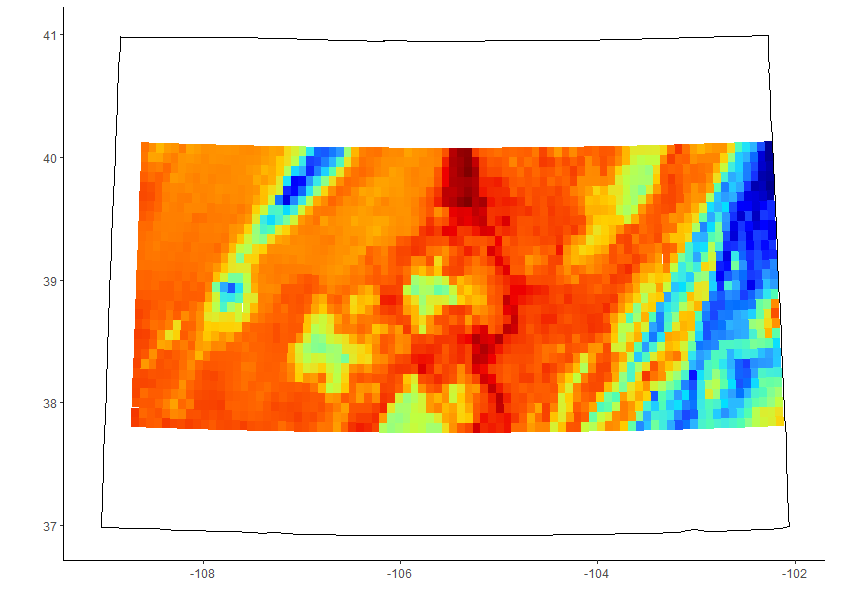}
\end{minipage} %
\begin{minipage}{0.49\linewidth}
\includegraphics[width=\linewidth]{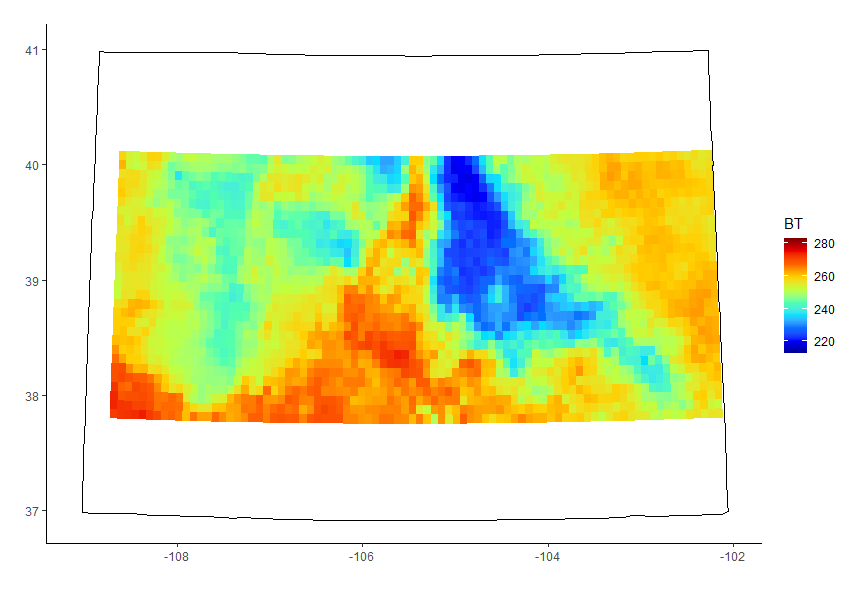}
\end{minipage} %
\vspace{-0.25cm}
\caption{Predicted brightness temperature fields at t = 196 and t = 325 obtained using corresponding smoothed wind estimates from the Space-Time Drift Model (STDM). The predictions have been made at a subset of spatial locations.}
\label{fig3.7}
\end{figure}

For this dataset, reference wind fields are not available. Hence, we compare the two methods based on Mean Squared Prediction Error (MSPE) while predicting standardized brightness temperature fields. To predict $Z$ at a spatial location $\bm{x}$ at time $t$,  we consider $\bm{Z}_D(\cdot, t - 1)$, the standardized data in $D(\bm{x}, t - 1)$ and estimated winds at time $t - 2$. This is because the estimated winds at time $t - 2$ uses data from times $t - 3, t - 2$ and $t - 1$. The predicted standardized temperature is calculated as the mean of the Gaussian conditional distribution of $Z(\bm{s}, t)$ given $\bm{Z}_D(\cdot, t - 1)$ under the stationary Mat\'ern space-time drift model with estimated drift parameter $\widehat{\bm{u}}(\bm{s}, t - 2)$. We predict standardized brightness temperature using the raw and smoothed versions of the wind estimates from both algorithms. We also consider a naive approach of predicting the brightness temperature fields, where the the data at time $t - 1$ is considered to be the predicted standardized brightness temperature fields at time $t$. We call it the persistence prediction and the performances of the two methods are assessed relative to the baseline MSPE. Table \ref{mytable4} compares the raw and smoothed wind estimates in terms of MSPE for 4 consecutive time points corresponding to the two data buffers. From Table \ref{mytable3}, we conclude that the proposed model outperforms the DMW Algorithm. We also see that smoothing the estimates results in more precise predicted brightness temperature fields and hence provides a better picture of the wind fields over Colorado. Figure \ref{fig3.7} provides maps of predicted brightness temperature at a subset of spatial locations and at $t = 196$ and $t = 325$ respectively using the smoothed estimated winds from STDM. It can be seen that our model captures the main features in the brightness temperature fields over Colorado. Our model also captures the wind movement over the region since we can see similar patterns of features across the region as compared to the ones tracked along in the original images (see Fig \ref{fig2a}). 

\begin{table}[bt]
\caption{\label{mytable4}Comparing Mean Squared Prediction Error based on prediction using the raw and smoothed wind estimates from the Space-Time Drift Model (STDM) and the Derived Motion Winds Algorithm (DMWA) at different time points. $\lambda$ in each case denotes the optimal smoothing parameter chosen using cross validation. All the methods have been compared against the baseline prediction.}
\centering
\resizebox{\textwidth}{!}{
\begin{tabular}{ c|cccc|cccc }
Method & $t = 196$ & $t = 197$ & $t = 198$ & $t = 199$ & $t = 325$ & $t = 326$ & $t = 327$ & $t = 328$ \\ \hline
STDM & \bf{0.329}  & \bf{0.316} & \bf{0.404} & \bf{0.732} & \bf{0.372} & \bf{0.588} & \bf{0.334} & \bf{0.684} \\
DMWA & 0.604 & 0.561 & 0.657 & 0.985 & 0.580 & 0.724 & 0.582 & 0.863 \\
Smoothed STDM $(\lambda = 2 \text{ km})$ & \bf{0.323} & \bf{0.300} & \bf{0.386} & \bf{0.716} & \bf{0.370} & \bf{0.580} & \bf{0.324} & \bf{0.619}\\
Smoothed DMWA $(\lambda = 8 \text{ km})$ & 0.486 & 0.454 & 0.553 & 0.899 & 0.505 & 0.664 & 0.493 & 0.776\\
Persistence & 0.596 & 0.568 & 0.614 & 1.137 & 0.684 & 1.204 & 0.613 & 1.180 \\ \hline
\end{tabular}}
\end{table}

\section{Discussions and Conclusions}

Wind is one of the most important atmospheric variables; it has large impact on local weather and hence, studying winds is essential. Wind data can be derived from high-resolution spatiotemporal data collected by geostationary satellites. These data are sequence of images over time and are used to derive atmospheric wind speed and direction. One algorithm that provides wind estimates is known as the Derived Motion Winds Algorithm. It takes as its input a sequence of images taken by the GOES-R series of the NOAA meteorological satellites and produces wind speed and direction. However, this algorithm does not quantify uncertainties. In this paper, we propose a spatiotemporal model to analyze satellite image data with the primary objective of estimating atmospheric wind speed and direction.

We propose local estimation of drift parameters. Developing a globally valid nonstationary space time drift model is an interesting problem that we have not pursued here. Following the basic idea of Nested Tracking, we propose a method to estimate the covariance parameters locally using maximum likelihood estimation. We smooth the raw estimated wind fields using a weighted Gaussian kernel, the weights being scaled by the inverse of the estimated variances of the estimates. Section 4 details an extensive simulation study that outlines conditions under which our model performs well. Based on our simulation study, we conclude that we have accurate wind estimates when the true spatial correlation range is small and the true temporal correlation range is high. The simulations also show that spatial smoothing provides better estimates of the wind fields. The simulations highlight a major drawback of the DMW Algorithm. Due to the design of the DMW Algorithm, the local DMW Algorithm estimates can only be half-integers and can only take values equal to the number of pixels the smaller target scene can move around in the larger search window. This brings us to perhaps the most important tuning parameter in the analysis, the target window size. We show that the window size is very crucial for both methods with large bias resulting from a large window and variance resulting form a small window.

We apply our method on brightness temperature data over Colorado, obtained from the GOES-15 satellite. While estimating winds, the window size has been chosen using cross validation. We provide estimated standard deviation maps, showing that our method is capable of quantifying uncertainties associated with the estimation. We also provide smoothed maps of estimated wind fields over Colorado. We predict brightness temperature fields using our model and compare the raw and smoothed estimates of wind fields based on Mean Squared Prediction Error (MSPE). We also compare the performance of our proposed method and the DMW Algorithm. We have shown that our method outperforms the DMW Algorithm with respect to MSPE. We argue that smoothing the estimated wind fields give more reliable wind estimates. We also see that we capture the main features in the brightness temperature maps through our prediction, including the drift across the region. 

One of the major challenges of our method is to apply it to data in real time. The main computational bottleneck is the time required for optimization over the covariance parameters. Because we use a local likelihood approach, the method is embarrassingly parallel across pixels and time points, and thus should scale well when used in production. The optimization can also be made faster using approximation. For example, the optimization algorithm could be initialized using the results of spatial or temporal neighbors, and the parameters could be updated using a one-step Fisher's scoring approximation. We would also like to explore more flexible methods for capturing the complicated wind motions, possibly a method where the window size is allowed to vary with location and time. This can ensure that our feature is always inside the frame of reference, resulting in more accurate estimates of the wind fields.

\section*{Appendix}
\noindent \textit{A. Standardization of data}: We standardize the brightness temperature data $Y(\bm{x}, t)$ as
$$
Z(\bm{x}, t) = \frac{Y(\bm{x}, t) - \widehat{\mu}(\bm{x})}{\widehat{\sigma}(\bm{x})}
$$
where $\widehat{\mu}(\bm{x})$ denotes the pixel-wise sample mean over time of $Y(\bm{x}, t)$ at location $\bm{x}$, and $\widehat{\sigma}(\bm{x})$ denotes the corresponding smoothed standard deviation. The standard deviation is smoothed using the Gaussian kernel 
$$\phi(\bm{x}|\bm{v}_l, \lambda) = \frac{1}{2\pi \lambda^2} \mbox{exp}\left( - \frac{||\bm{x} - \bm{v}_l||^2}{2\lambda^2}\right),$$
where $\lambda > 0$ is the kernel bandwidth and $\bm{v}_1, \ldots, \bm{v}_L \in \mathbb{R}^2$ denote the spatial locations over the entire region considered. The smoothing weights are defined as  
\begin{equation*} 
w_l(\bm{x}) = \frac{\phi(\bm{x}|\bm{v}_l, \lambda)}{\sum_{j = 1}^L\phi(\bm{x}|\bm{v}_j, \lambda)}.
\end{equation*}
That is, for standardizing the data we use
\begin{equation*}
\widehat{\sigma}(\bm{x}) = \sum_{l = 1}^L \tilde{\sigma}(\bm{v}_l) w_l(\bm{x}),
\end{equation*}
where $\tilde{\sigma}(\bm{v}_l)$ is the sample standard deviation at location $\bm{v}_l$.

\medskip

\noindent \textit{B. Smoothing the estimates}: Let $\widehat{\bm{u}}(\bm{x}, t) = \left( \widehat{u}(\bm{x}, t), \widehat{v}(\bm{x}, t)\right)$ denote the estimated wind vector in $D(\bm{x}, t)$ using the proposed method. Also let the corresponding estimated variances be denoted by $\widehat{d}_u(\bm{x}, t)$ and $\widehat{d}_v(\bm{x}, t)$ where the suffixes `u' and `v' refer to the u- and v- component of the estimated wind vector. We smooth each component of the estimates using a weighted Gaussian smoothing filter, the weights being equal to the inverses of the corresponding variance estimates. Let $\widehat{\bm{u}}^{(s)}(\bm{x}, t) = \left( \widehat{u}^{(s)}(\bm{x}, t), \widehat{v}^{(s)}(\bm{x}, t)\right)$ denote the smoothed wind estimates where,
\begin{align*}
\widehat{u}^{(s)}(\bm{x}, t) &= \sum_{l = 1}^L \widehat{u}(\bm{v}_l, t) w_l^u(\bm{x}, t), \hspace{0.35cm} w_l^u(\bm{x}, t) = \frac{\Big\{ \phi(\bm{x}|\bm{v}_l, \lambda)/\widehat{d}_u(\bm{v}_l, t)\Big\} }{\sum_{j = 1}^L\Big\{ \phi(\bm{x}|\bm{v}_j, \lambda)/\widehat{d}_u(\bm{v}_j, t)\Big\} }
\end{align*}
\begin{align*}
\widehat{v}^{(s)}(\bm{x}, t) &= \sum_{l = 1}^L \widehat{v}(\bm{v}_l, t) w_l^v(\bm{x}, t), \hspace{0.35cm} w_l^v(\bm{x}, t) = \frac{\Big\{ \phi(\bm{x}|\bm{v}_l, \lambda)/\widehat{d}_v(\bm{v}_l, t)\Big\} }{\sum_{j = 1}^L\Big\{ \phi(\bm{x}|\bm{v}_j, \lambda)/\widehat{d}_v(\bm{v}_j, t)\Big\} }
\end{align*}

\bibliographystyle{agsm}

\bibliography{newbib}
\end{document}